\documentclass[sn-mathphys,Numbered]{sn-jnl}

\usepackage{graphicx}%
\usepackage{multirow}%
\usepackage{amsmath,amsfonts,amssymb,array,tabularx,enumerate}%
\usepackage{amsthm}%
\usepackage{mathrsfs}%
\usepackage[title]{appendix}%
\usepackage{xcolor}%
\usepackage{textcomp}%
\usepackage{manyfoot}%
\usepackage{booktabs}%
\usepackage{algorithm}%
\usepackage{algorithmicx}%
\usepackage{algpseudocode}%
\usepackage{listings}
\usepackage{float}
\usepackage{caption}
\usepackage{subcaption}%

\begin{document}

\title[The EE-Classifier: A classification method for functional data based on extremality indexes]{The EE-Classifier: A classification method for functional data based on extremality indexes}

\author*[1]{\fnm{Catalina} \sur{Lesmes}}\email{clesmes@eafit.edu.co}
\author[1]{\fnm{Francisco} \sur{Zuluaga}}\email{fzuluag2@eafit.edu.co}
\author[2]{\fnm{Henry} \sur{Laniado}}\email{hlaniado@unicauca.edu.co}
\author[1]{\fnm{Andres} \sur{Gomez}}\email{agomeza10@eafit.edu.co}
\author[1]{\fnm{Andrea} \sur{Carvajal}}\email{acarvajalm@eafit.edu.co}
\affil*[1]{\orgdiv{School of Applied Sciences and Engineering}, \orgname{Universidad EAFIT}, \orgaddress{\city{Medellin}, \country{Colombia}}}
\affil[2]{\orgdiv{Mathematics Department}, \orgname{Universidad del Cauca}, \orgaddress{ \city{Popayan}, \country{Colombia}}}

\abstract{Functional data analysis has gained significant attention due to its wide applicability. This research explores the extension of statistical analysis methods for functional data, with a primary focus on supervised classification techniques. It provides a review on the existing depth-based methods used in functional data samples. Building on this foundation, it introduces an extremality-based approach, which takes the modified epigraph and hypograph indexes properties as classification techniques. To demonstrate the effectiveness of the classifier, it is applied to both real-world and synthetic data sets. The results show its efficacy in accurately classifying functional data. Additionally, the classifier is used to analyze the fluctuations in the S\&P 500 stock value. This research contributes to the field of functional data analysis by introducing a new extremality-based classifier. The successful application to various data sets shows its potential for supervised classification tasks and provides valuable insights into financial data analysis.}

\maketitle
\section{Introduction}\label{sec1}

Data processing and analysis has become essential for understanding various phenomena, whether found in research, business, politics, or other domains, it provides interesting insights and guide to decision making. As we adapt to an evolving world based on data, our approach to data analysis must evolve as well. This has led to the emergence of functional data analysis, which enables us to model data from an innovative perspective, in this particular case, as a function. 

The concept of functional data, first coned by \cite{bib1} has been explored over the years, providing interesting research and findings. As this concept develops, the demand for statistical techniques customized to its distinct characteristics also grows. Traditional statistical techniques are extended to analyze this new data form. Approaches like mean and covariance computations have been expanded to serve as descriptive tools for functional data \cite{bib2}. Additionally, more advanced techniques such as classification \cite{bib3}, clustering \cite{bib4}, principal component analysis \cite{bib5}, homogeinity tests \cite{bib6} and so on, have been adapted to suit functional data analysis.

This article is centered around the classification aspect of functional data analysis. Major advances on classification have been made using functional depth \cite{bib7}, the most successful classifiers, the DD-classifier \cite{bib8} and the $DD^G$-classifier \cite{bib9}, provide a great example of how functional depth is applied successfully for classification. For this research, we plan on extending these classification methods to incorporate function extremality indexes.

Extremality for functional data \cite{bib10} provides a natural ordering for sample curves by introducing the definitions of the hypograph and epigraph indexes. These concepts have been used to develop different methodologies in the functional data context, such as the outliergram \cite{bib11}, the functional boxplot \cite{bib12}, rank tests \cite{bib13} and clustering \cite{bib14}. Giving such promising results, they inspire to explore different methodologies, such as classification.

Therefore, the purpose of this research is to explore how the methodology of existing depth-based classifiers can be extended to extremality-based methods. It shows the extension of the methodology to extremality concepts and introduces the EE-plot and EE-classifier. Alongside, it provides illustration of their functionality by creating experiments using synthetic and real data provided by R libraries. Furthermore, a real application of the classifier will used to predict the S\&P stock value fluctuation.

The subsequent sections of this article are structured as follows: Section \ref{sec2} introduces fundamental preliminary concepts related to the depth-based classifiers and function extremality: epigraph and hypograph. In Section \ref{sec3}, the EE-plot and EE-classifier are described. Section \ref{sec4} presents results obtained using the EE-classifier with simulated and real datasets, followed by an application to S\&P stock value changes in Section \ref{sec5}. Finally, Sections \ref{sec6} offer concluding remarks followed by acknowledgement and references.

\vspace{5mm}

\section{Preliminary Concepts}\label{sec2}

It remains helpful to take a closer look at some essential concepts that form the basis of our research.  The proposed method for classifying functional data is built upon the principles of the DD-classifier \cite{bib8} and the $DD^G$-classifier \cite{bib9}. To gain a better grasp of our approach, these concepts will be defined. Additionally, understanding the concepts of function epigraph and hypograph is essential to fully comprehend the new ideas introduced in this article.

\subsection{DD-classifier $\&$ $DD^{G}$-classifier}

The DD-classifier \cite{bib8} and $DD^G$-classifier \cite{bib9} are founded upon the concept of the data-depth plot. The DD-plot \cite{bib20} illustrates the depth values of the sample in relation to two distributions. This transition takes the data from the functional space to the  depth-depth space. For the DD-plot you could use any of the implemented depths, some of the most common functional depths are reviewed.

\begin{itemize}
    \item Fraiman and Muniz (FM) depth: Proposed by \cite{bib21}, provides the depth measurements for the $i-th$ data point. Given a sample $x_1,...,x_N$ of functions defined on the interval $[0,T]$ and $S_t$ = $(x_1(t), ... , x_N(t))$ the values of those functions on a given $t \in [0,T]$. Let the empirical distribution of the sample $S_t$ denoted by $F_{N,t}$ and let $Z_i(t)$ a univariate depth of $x_i(t)$ in this sample. The FM depth for the $i-th$ data point is given by the following integral: 
    \begin{equation}
        FM_i = \int_{0}^{T} Z_i(t)dt
    \label{eq:0.0}
    \end{equation}

    For the original definition $Z_i(t)$ = $1 - |1/2 - F_{N,t}(x_i(t))|$, but this can change according the different univariate depths to consider, which could be the Tukey depth, the Simplicial depth, Likelihood depth and the Mahalanobis depth.\\
    
    \item h-mode (hM) depth: This depth, as seen in \cite{bib22} can be described as a generalization for the Likelihood depth, where you could measure how surrounded one curve is with respect to the others. The population h-mode depth of a datum $x_0$ is given by:

    \begin{equation}
        f_h(x_0) = E[K(m(x_0, X)/h)]
        \label{eq:0.1}
    \end{equation}
    
    where $X$ is the random element describing the population, $m$ is the suitable metric, $K(t)$ is an asymmetric kernel and $h$ is the bandwidth parameter. Given a random sample $x_1, ..., x_N$ of X, the empirical h-mode depth is defined as: 

    \begin{equation}
        \hat{f}_h(x_0) = N^{-1} \sum_{i=1}^N K(m(x_0, x_i)/h) \\
        \label{eq:0.2}
    \end{equation}
    
    \item Random projections (RP) depth: This method proposed by \cite{bib22} uses the univariate half space depth and summarizes the depths of the projection through the mean. The definition for the depth is then given by: 

    \begin{equation}
        RP(x) = R^-1 \sum_{r=1}^RD_{a_r}(x)
        \label{eq:0.3}
    \end{equation}
    
    where $D{a_r}(x)$ is the depth associated with the $r-th$ projection and using $R=50$ as the default choice. If we wanted to extend this method to the multivariate case we could also construct a weighted depth from the depth of each component. 

\end{itemize}

\begin{figure}[h]
    \centering
    
    \begin{subfigure}{0.45\textwidth}
        \centering
        \includegraphics[width=\linewidth]{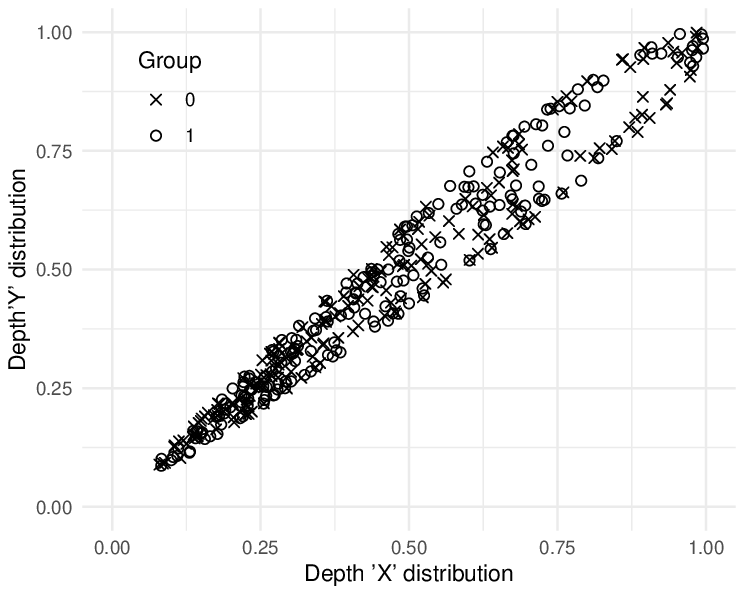} 
        \caption{Similar distributions}
    \end{subfigure}
    \hfill
    \begin{subfigure}{0.45\textwidth}
        \centering
        \includegraphics[width=\linewidth]{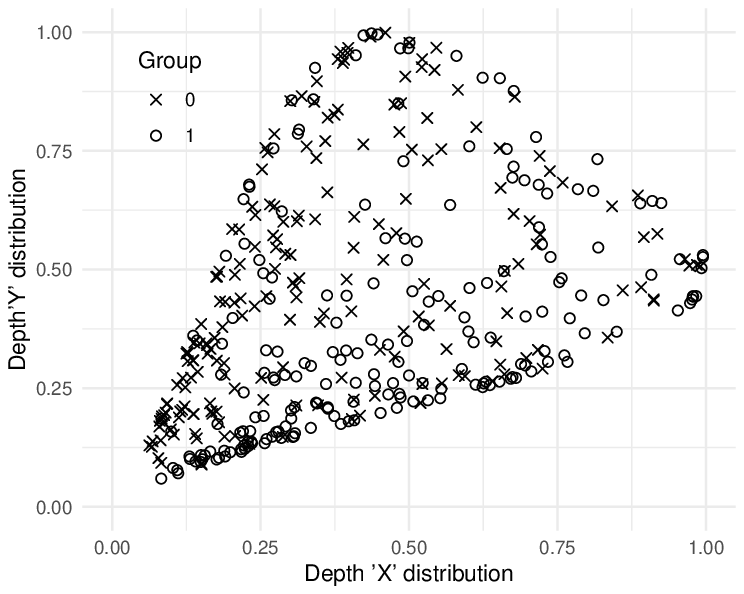} 
        \caption{Different distributions}
    \end{subfigure}
    
    \caption{DD-plot illustration with different $X$ and $Y$ distributions}
    \label{fig:ddplot}
\end{figure}

The formal definition of the DD-plot is given by \cite{bib20}. Let us consider two distinct distributions, $F$ and $G$ defined on $\mathbb{R}^d \to \mathbb{R}$. Given an invariant depth measure denoted as $D(\cdot)$, $DD(F,G)$ is defined as follows:

\begin{equation}
DD(F,G) = \{(D_{F}(x),D_{G}(x)), \forall x \in \mathbb{R}^d \}
\label{eq:1}
\end{equation}

\vspace{0.2cm}

When the underlying distribution $F$ is unknown, with a given dataset  $X_{1}, ... , X_{n}$, once could determine whether $F$ is some specified distribution, for example $G$, by examining the following DD-plot:

\begin{equation}
DD(F_{n},G) = \{(D_{F_{n}}(x),D_{G}(x)), \forall x \in \{X_1, . . . , X_n\} \}
\label{eq:2}
\end{equation}

\vspace{0.2cm}

If $F$ and $G$ are the population distributions for the samples $\mathcal{X} = \{X_1, . . . , X_n\}$ and $\mathcal{Y} = \{Y_1, . . . , Y_m\}$ then the DD-plot below can be used to determine whether or not the two distributions are identical:

\begin{equation}
DD(F_{n},G_{m}) = \{(D_{F_{n}}(x),D_{G_{m}}(x)), x \in \{\mathcal{X}  \cup \mathcal{Y} \} \}
\label{eq:3}
\end{equation}

To illustrate the DD-plot let us take a look at Figure \ref{fig:ddplot}, where two different distributions with 200 samples each were generated using the \emph{mvrnorm} function from the \emph{MASS} R package. This function produces samples from a specified multivariate normal distribution. Figure \ref{fig:ddplot} shows the DD-plot using the Mahalanobis depth, \ref{fig:ddplot}(a) with two $ N (0,0) $ distributions and Figure \ref{fig:ddplot}(b) with one $ N (0,0) $ distribution and another $ N (0,1) $ distribution.

The research presented in \cite{bib20} provides valuable insights into the characterization of different curves using the DD-plot. In instances where the two distributions are similar, the DD-plot should be concentrated along the diagonal line, as exemplified in Figure \ref{fig:ddplot}(a). Notable patterns or deviations from this line, as seen in Figure \ref{fig:ddplot}(b), indicate specific differences between distributions F and G. These distinctions may arise from a location shift, manifesting as a heart-shaped DD-plot; a scale difference, resulting in an arched DD-plot above the diagonal line; skewness differences, producing an asymmetric arch on the DD-plot; or kurtosis differences, leading to a shift in the lower part of the DD-plot to one side of the diagonal line. This visual tool offers a comprehensive understanding of the diverse characteristics that can be characterized through the DD-plot.

This characterization from the DD-plot creates a space where one could use different classification methods to specify whether the two distributions have similar characteristics. That is where the DD-classifier and the $DD^G$-classifier use their methodologies to determine the similarities between distributions. 

The distinction between these two classifiers remains as the way of how both classifiers create classification rules based on the DD-plot. The DD-classifier transforms the data from the functional space to the depth-depth space and aims to find the optimal function that separates the groups based on the minimal classification error. Conversely, the $DD^G$-classifier provides classification based on traditional classification methods such as Linear Discriminant Analysis, Quadratic Discriminant Analysis, k-Nearest Neighbors, and others. Besides, unlike the $DD$-classifier, which allows only two classification groups at a time, the $DD^G$-Classifier extends this capability to more than two groups.

\subsection{Epigraph and Hypograph Indexes}
The concepts of extremality include the epigraph and hypograph of a function. The hypograph of a function shows the points situated either directly on or below its graph. On the contrary, the epigraph shows points either directly on or above the graph of the function. In a formal sense, based on \cite{bib15}, let $C(\mathcal{I})$ denote the space of continuous functions defined within a compact interval $\mathcal{I}$. Let us consider a stochastic process $X$ with sample paths in $C(\mathcal{I})$ and distributed according to $P$. Suppose $x_1(t), x_2(t), ... , x_n(t)$ represents a sample of curves from $P$. We will denote the graph of the function $x$ in $C(\mathcal{I})$ as $G(x)$, thus $G(x) = {(t,x(t)), t \in \mathcal{I} }$. Following this, the definitions for the hypograph and the epigraph of a function $x$ in $C(\mathcal{I})$ can be expressed as:

\begin{equation}
\begin{cases}
\begin{split}
hyp(x) = {(t,y) \in \mathcal{I} \times \mathbb{R} : y > x(t) }\ \\
epi(x) = {(t,y) \in \mathcal{I} \times \mathbb{R} : y \leq x(t)} \
\end{split}
\end{cases}
\label{eq:4}
\end{equation}

\vspace{0.2cm}

The hypograph and epigraph indexes were introduced \cite{bib10} to provide an order for the curves in functional data. These indexes  provide a proportion of functions within the sample whose graphs fall within the hypograph or epigraph of function $x$. This quantification is obtained by subtracting the proportion of curves in the sample that lie below or above $x$ from one. Formally, these can be defined as follows:

 \begin{equation}
 \begin{cases}
\begin{split}   
     HI(x) & = \frac{1}{n} \sum_{i=1}^n I( G(x_i) \subset hyp(x)  ) \\
     & = \frac{1}{n} \sum_{i=1}^n I( x_i(t) \leq x(t), t \in \mathcal{I}) \\
     EI(x) & = \frac{1}{n} \sum_{i=1}^n I( G(x_i) \subset epi(x) ) \\
      & = \frac{1}{n} \sum_{i=1}^n I( x_i(t) \geq x(t), t \in \mathcal{I}) 
    \end{split}
 \end{cases}
 \label{eq:5}
 \end{equation}

\vspace{0.2cm}

where $I(A)$ is the indicator function of the set A. The concepts of hypograph and epigraph indexes may encounter limitations in scenarios where curve classifications involve closely or widely spaced curves \cite{bib11}. Consequently, modified versions of these indices were also introduced \cite{bib15} to address such situations:

\begin{equation}
\begin{cases}
\begin{split} 
    MHI(x) = \frac{1}{n \lambda(\mathcal{I})} \sum_{i=1}^n \lambda \{t \in \mathcal{I}:  x_i(t) \leq x(t) \} \\
    MEI(x) = \frac{1}{n \lambda(\mathcal{I})} \sum_{i=1}^n \lambda \{t \in \mathcal{I}: x_i(t) \geq x(t) \}
\end{split}
\end{cases}
 \label{eq:6}
\end{equation}

\vspace{0.2cm}

where $\lambda$ stands for the Lebesgue measure on $\mathbb{R}$. One important characteristic given by MHI and MEI, described in \cite{bib14}, is that they comply the following relation $MHI(x) - MEI(x)= \frac{1}{n}$, concluding that they are linearly dependent. 

\section{EE-plot \& EE-classifier}\label{sec3}
As described in Section \ref{sec2}, the modified epigraph and hypograph indexes, show characteristics for curve samples, giving them an order, even when the curves are closely or widely spaced. This provides information about distribution's similarities if one were to compare them. Visualization tools to provide comparison between curve samples have been used before, such as in \cite{bib13} where they provide homogeneity testing between samples and in \cite{bib14} where clustering techniques are applied to separate them. Therefore, we aim to explore a visualization tool which we name as Extremality-Extremality (EE) plot as a method to apply classification techniques and introduce a new classification method named as the Extremality-Extremality (EE) classifier. 

\subsection{EE-plot}
The EE-plot illustrates the modified epigraph or hypograph indexes of the sample combined by two distributions. Following a similar methodology as in equation \ref{eq:1}, consider $F$ and $G$ two different distributions in $\mathbb{R}^d \to \mathbb{R}$ and let $MEI(\cdot)$ the modified epigraph index and $MHI(\cdot)$ the modified hypograph index, then EE(F,G) can be defined in two ways $EE_{H}(F,G)$ for when the modified hypograph index is used and $EE_{E}(F,G)$ for when the modified epigraph index is used, the definitions are given as:

\begin{equation}
    \begin{cases}
    EE_{H}(F,G) = \{(MHI_{F}(x), MHI_{G}(x)), \ \forall x \in \mathbb{R}^d \}\\
    EE_{E}(F,G) = \{(MEI_{F}(x), MEI_{G}(x)), \ \forall x \in \mathbb{R}^d \}
    \end{cases} 
 \label{eq:6}
\end{equation}

The EE-plot establishes a new space derived from the characteristics of functional data, therefore providing a transformation from the functional space into $\mathbb{R}^2$ for our data. Let us suppose we have two different population distributions for the samples $\mathcal{X} = \{X_1, . . . , X_n\}$ and $\mathcal{Y} = \{Y_1, . . . , Y_n\}$, as in the equation \ref{eq:3} we could define the EE-plot based on the union defined from $\mathcal{X}$ and $\mathcal{Y}$ as: 

\begin{equation}
\begin{cases}
EE_{H}(F_{n},G_{n}) = \{(MHI_{F_{n}}(x),MHI_{G_{n}}(x)), x \in \{\mathcal{X} \cup \mathcal{Y}\} \}\\
EE_{E}(F_{n},G_{n}) = \{(MEI_{F_{n}}(x),MEI_{G_{n}}(x)), x \in \{\mathcal{X} \cup \mathcal{Y}\} \}
\end{cases}
 \label{eq:7}
\end{equation}

\vspace{0.2cm}

\begin{figure}[h]
\centering
\captionsetup[subfigure]{justification=centering}
\begin{subfigure}[t]{.32\textwidth}
  \centering
  \includegraphics[width=.8\linewidth, height=10cm, keepaspectratio=true]{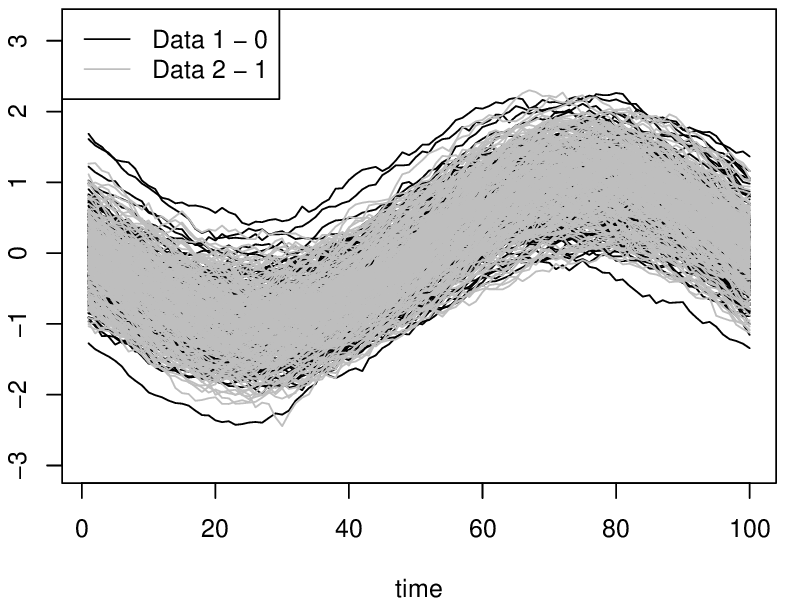}
  \caption{Experiment 1: \\Distribution behavior}
\end{subfigure}%
\hfill
\begin{subfigure}[t]{.32\textwidth}
  \centering
  \includegraphics[width=.8\linewidth, height=4cm, keepaspectratio=true]{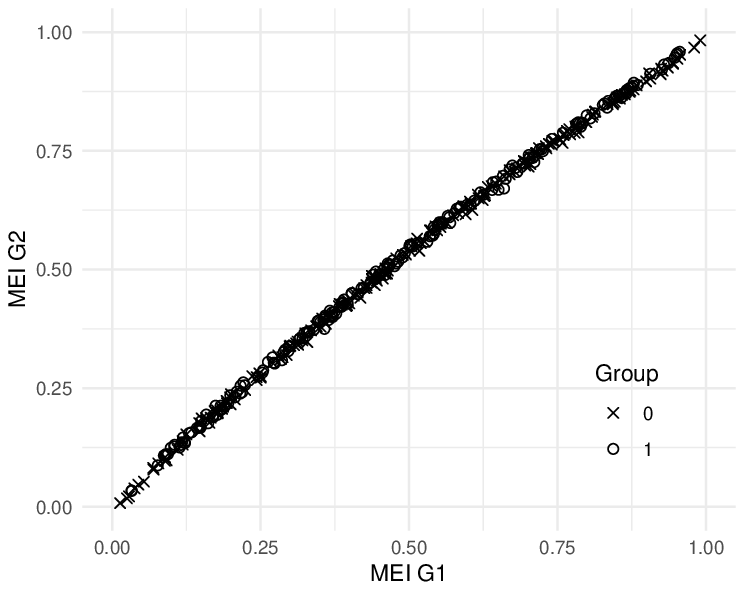}
  \caption{Experiment 1: \\$EE_{E}$-plot}
\end{subfigure}%
\hfill
\begin{subfigure}[t]{.32\textwidth}
  \centering
  \includegraphics[width=.8\linewidth, height=4cm, keepaspectratio=true]{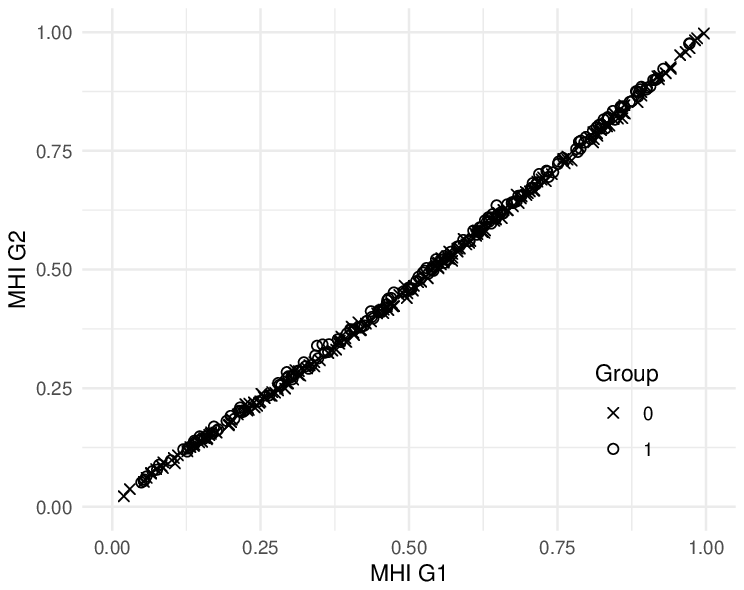}
  \caption{Experiment 1: \\$EE_{H}$-plot}
\end{subfigure}

\begin{subfigure}[t]{.32\textwidth}
  \centering
  \includegraphics[width=.8\linewidth, height=10cm, keepaspectratio=true]{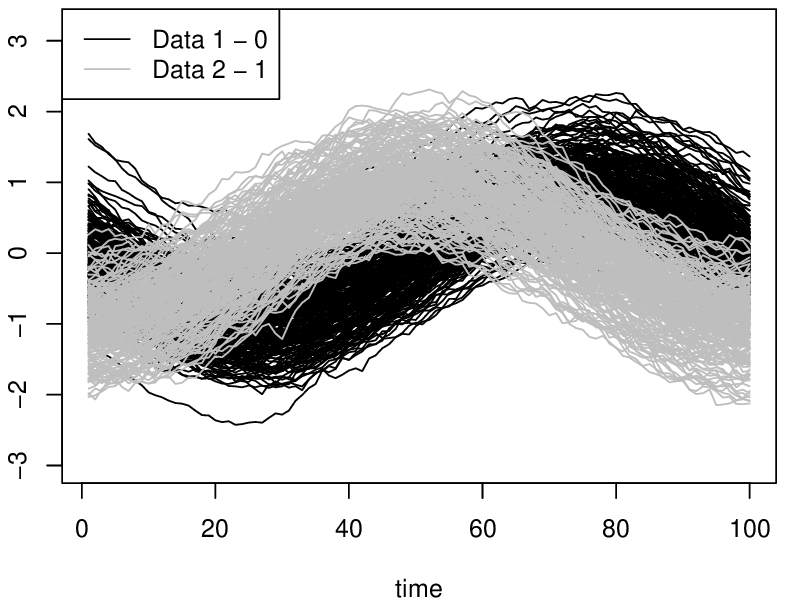}
  \caption{Experiment 2: \\Distribution behavior}
\end{subfigure}%
\hfill
\begin{subfigure}[t]{.32\textwidth}
  \centering
  \includegraphics[width=.8\linewidth, height=4cm, keepaspectratio=true]{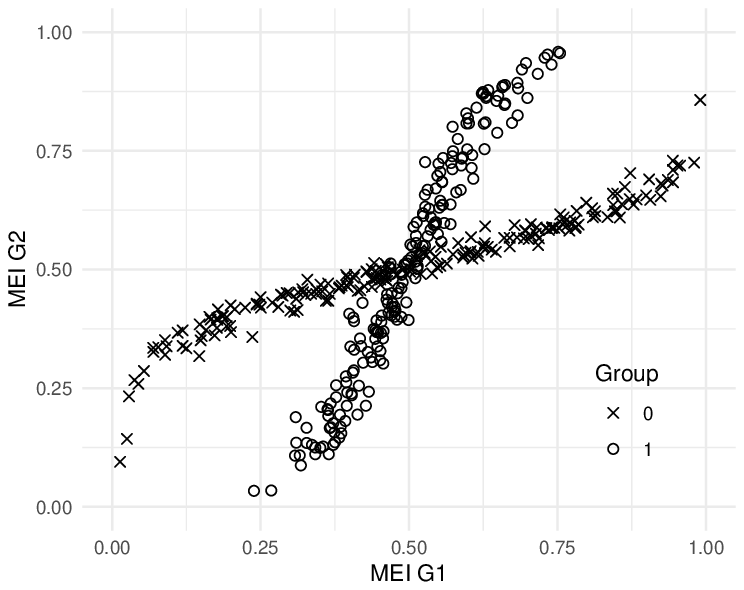}
  \caption{Experiment 2: \\$EE_{E}$-plot}
\end{subfigure}%
\hfill
\begin{subfigure}[t]{.32\textwidth}
  \centering
  \includegraphics[width=.8\linewidth, height=4cm, keepaspectratio=true]{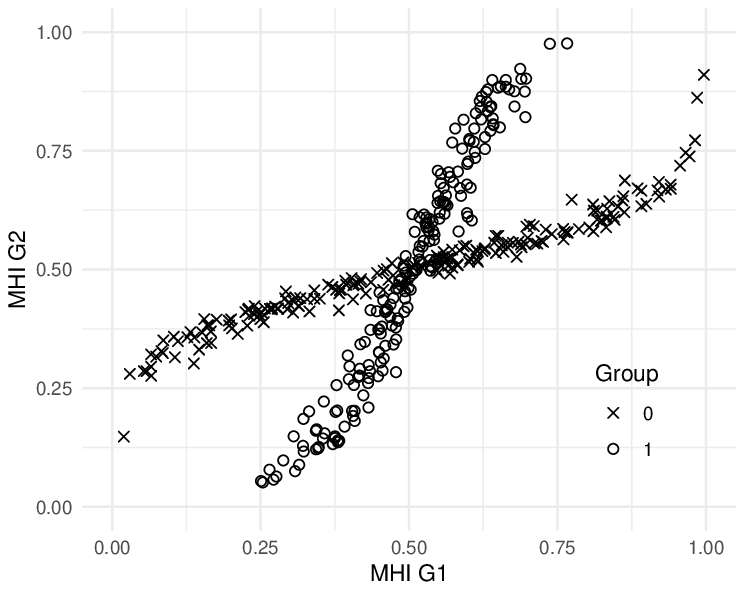}
  \caption{Experiment 2: \\$EE_{H}$-plot}
\end{subfigure}

\begin{subfigure}[t]{.32\textwidth}
  \centering
  \includegraphics[width=.8\linewidth, height=10cm, keepaspectratio=true]{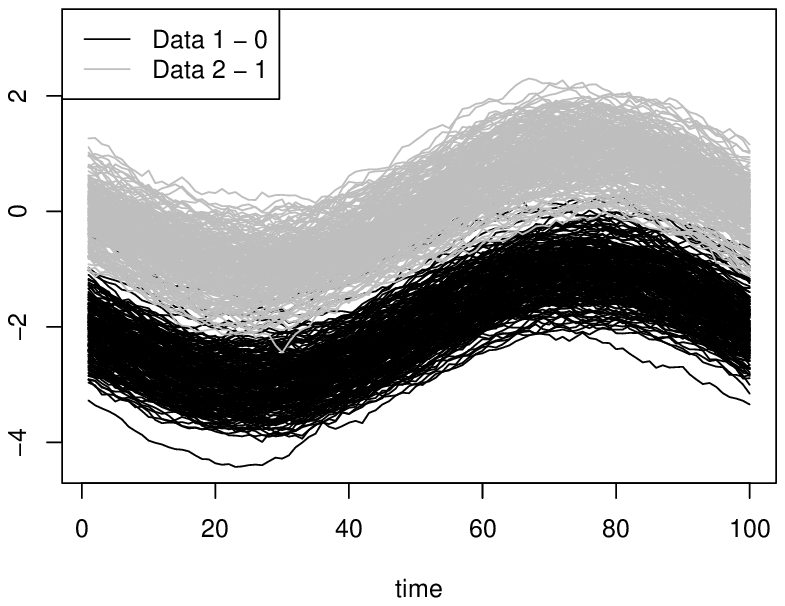}
  \caption{Experiment 3: \\Distribution behavior}
\end{subfigure}%
\hfill
\begin{subfigure}[t]{.32\textwidth}
  \centering
  \includegraphics[width=.8\linewidth, height=4cm, keepaspectratio=true]{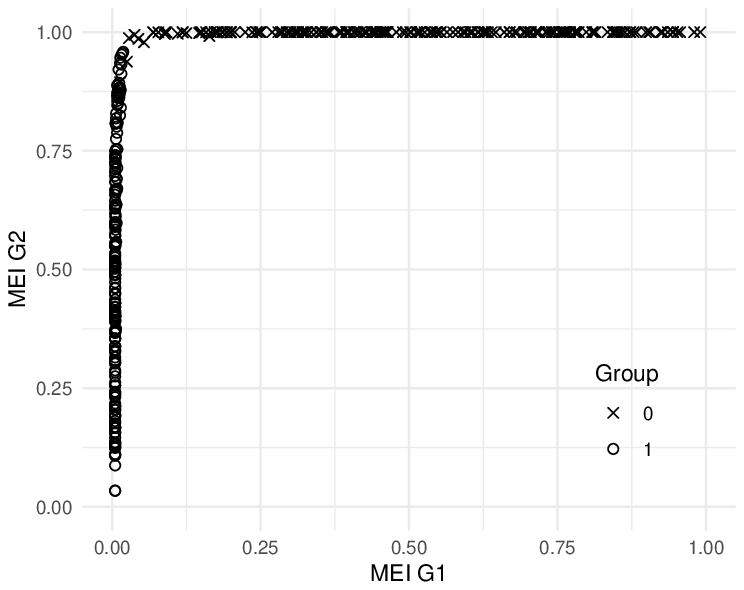}
  \caption{Experiment 3: \\$EE_{E}$-plot}
\end{subfigure}%
\hfill
\begin{subfigure}[t]{.32\textwidth}
  \centering
  \includegraphics[width=.8\linewidth, height=4cm, keepaspectratio=true]{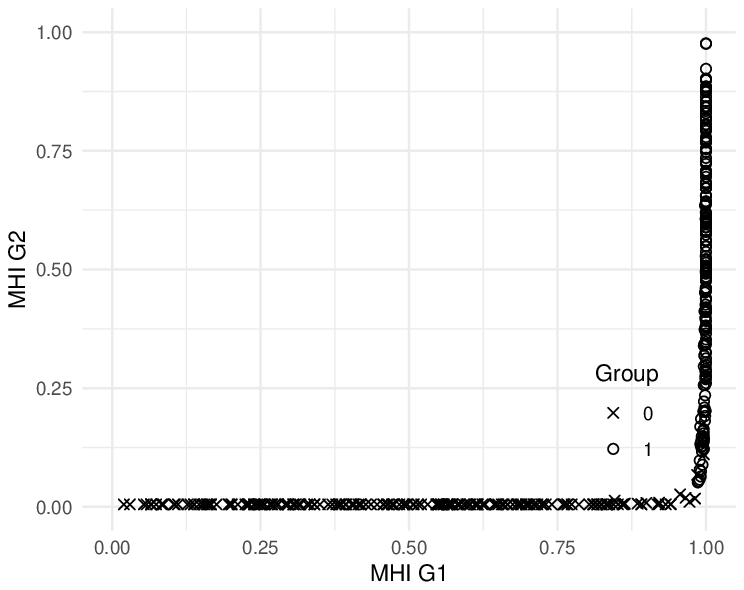}
  \caption{Experiment 3: \\$EE_{H}$-plot}
\end{subfigure}

\caption{Illustration of EE-plots given by two distributions from different experiments}
\label{fig:1_experiments}
\end{figure}

Illustrated in Figure \ref{fig:1_experiments}, the $EE_{H}$ and $EE_{E}$ application is demonstrated for two distinct distributions. Notably, Figures \ref{fig:1_experiments}(b), \ref{fig:1_experiments}(c), \ref{fig:1_experiments}(e), and \ref{fig:1_experiments}(f) depict how the $EE_{H}$ and $EE_{E}$ exhibit a mirrored relationship, highlighting the underlying behavior given by the relation $MHI(x) - MEI(x)= \frac{1}{n}$.

\begin{figure}[h]
\centering
\captionsetup[subfigure]{justification=centering}
\begin{subfigure}{.34\textwidth}
  \centering
  \includegraphics[width=.8\linewidth, height=10cm,keepaspectratio=true]{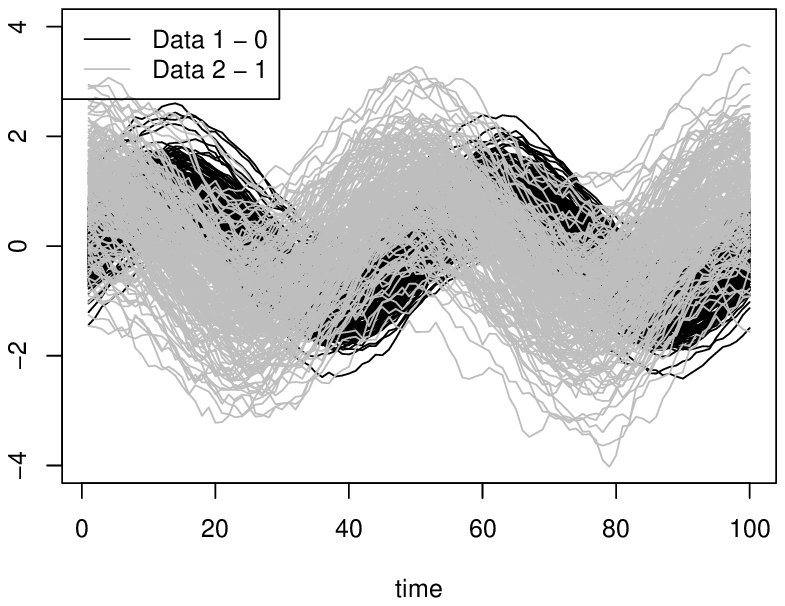}
  \caption{Experiment 4: \\Distribution behaviour}
\end{subfigure}%
\begin{subfigure}{.33\textwidth}
  \centering
  \includegraphics[width=.8\linewidth, height=4cm,keepaspectratio=true]{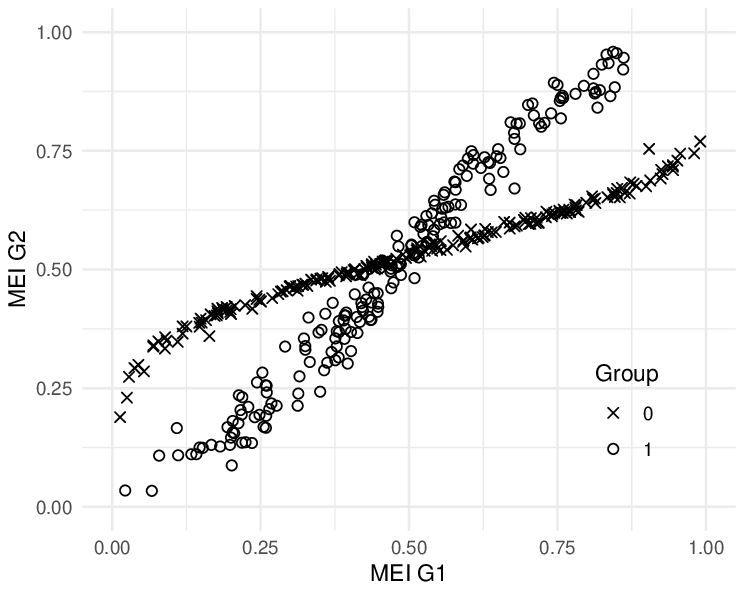}
  \caption{Experiment 4:\\$EE_{E}$-plot}
\end{subfigure}%
\begin{subfigure}{.33\textwidth}
  \centering
  \includegraphics[width=.8\linewidth, height=4cm,keepaspectratio=true]{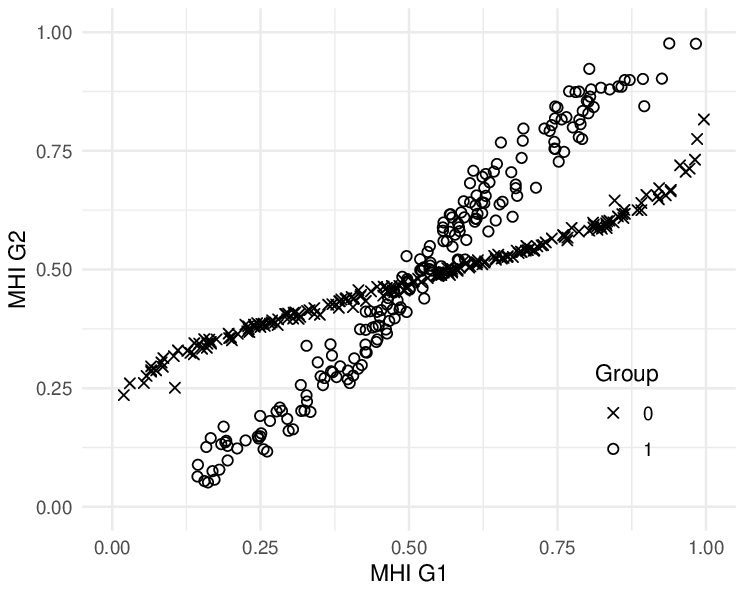}
  \caption{Experiment 4: \\$EE_{H}$-plot}
\end{subfigure}
\hfill
\begin{subfigure}{.34\textwidth}
  \centering
  \includegraphics[width=.8\linewidth, height=10cm,keepaspectratio=true]{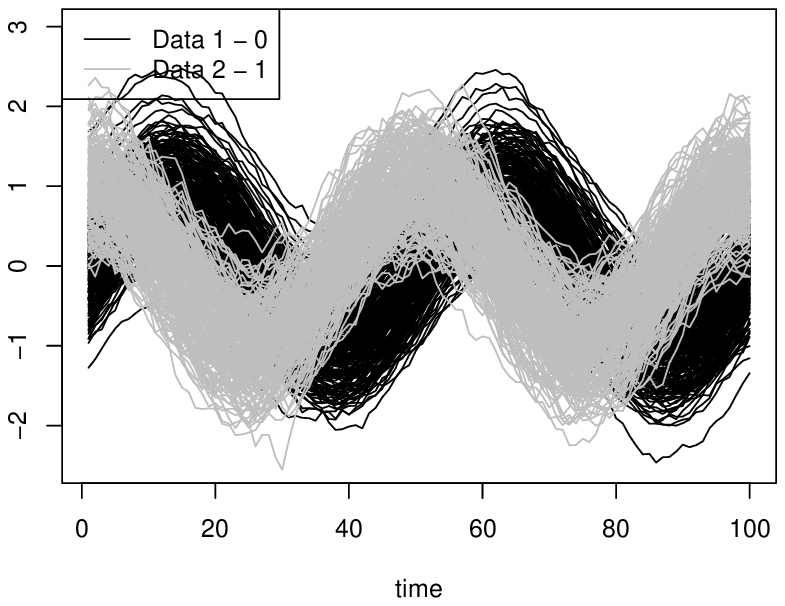}
  \caption{Experiment 5: \\Distribution behaviour}
\end{subfigure}%
\begin{subfigure}{.33\textwidth}
  \centering
  \includegraphics[width=.8\linewidth, height=4cm,keepaspectratio=true]{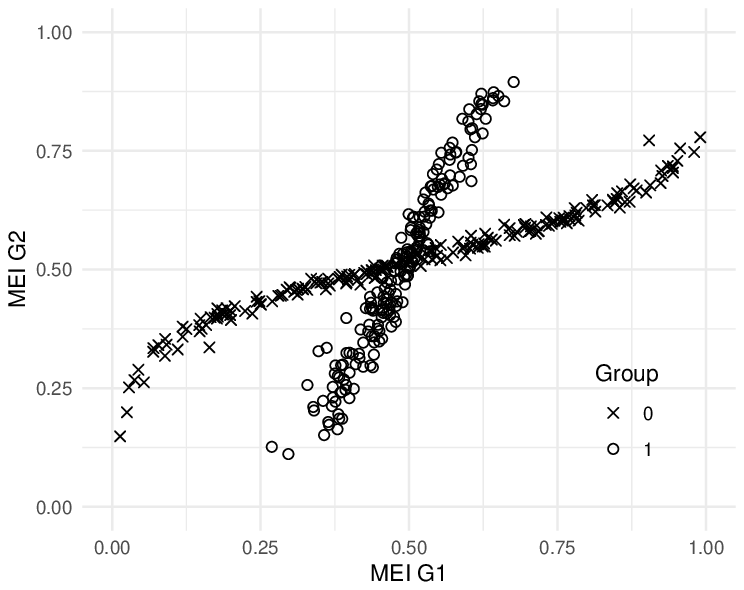}
  \caption{Experiment 5: \\$EE_{E}$-plot}
\end{subfigure}%
\begin{subfigure}{.33\textwidth}
  \centering
  \includegraphics[width=.8\linewidth, height=4cm,keepaspectratio=true]{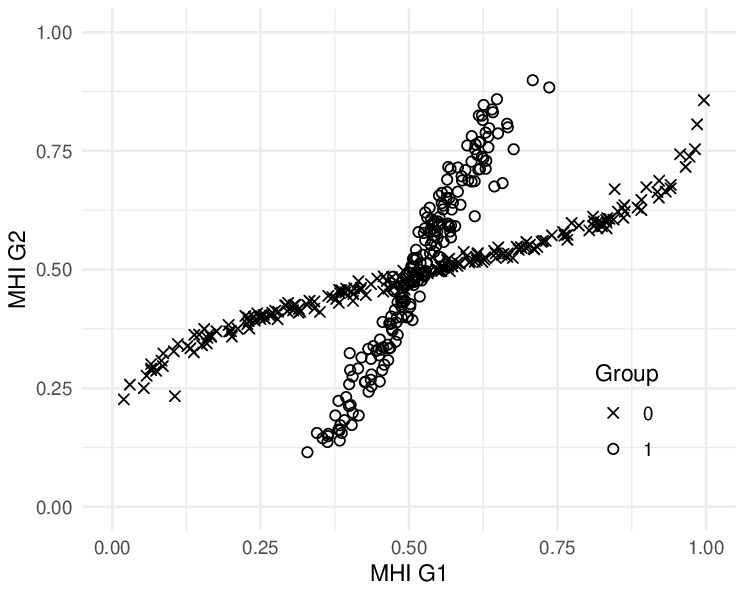}
  \caption{Experiment 5: \\$EE_{H}$-plot}
\end{subfigure}
\hfill
\begin{subfigure}{.34\textwidth}
  \centering
  \includegraphics[width=.8\linewidth, height=10cm,keepaspectratio=true]{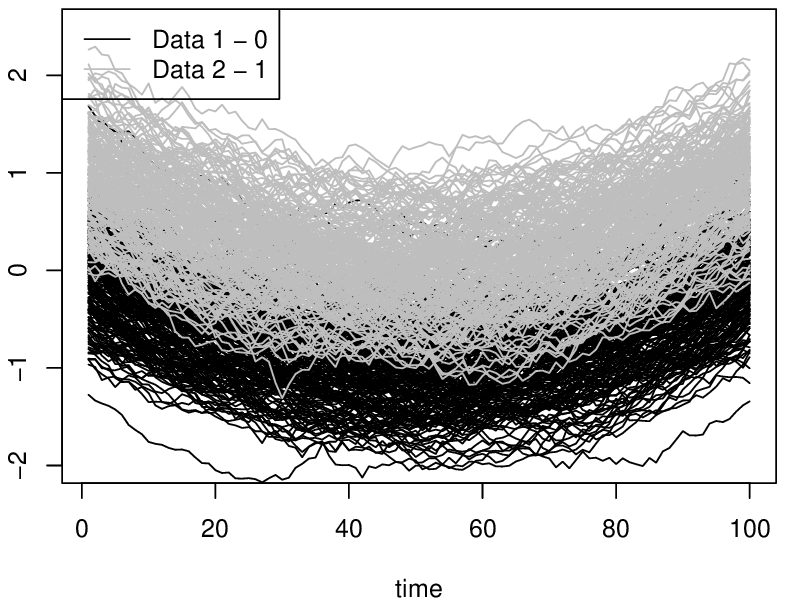}
  \caption{Experiment 6: \\Distribution behaviour}
\end{subfigure}%
\begin{subfigure}{.33\textwidth}
  \centering
  \includegraphics[width=.8\linewidth, height=4cm,keepaspectratio=true]{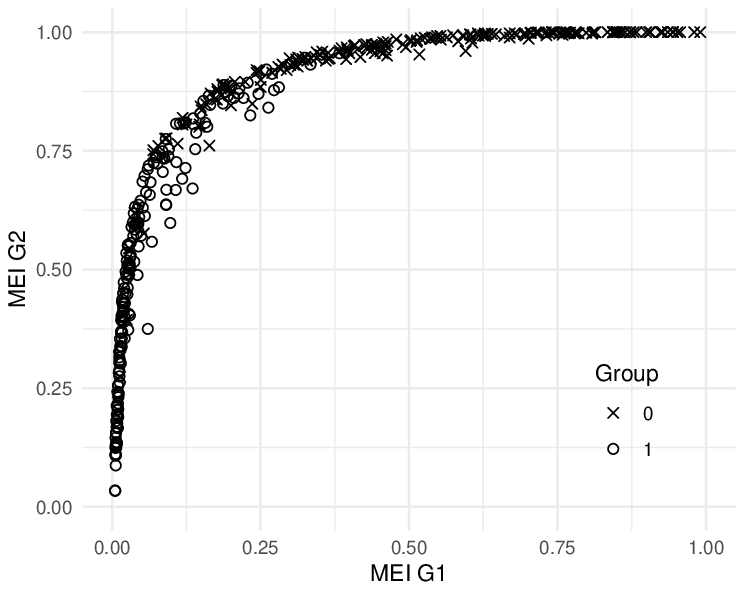}
  \caption{Experiment 6: \\$EE_{E}$-plot}
\end{subfigure}%
\begin{subfigure}{.33\textwidth}
  \centering
  \includegraphics[width=.8\linewidth, height=4cm,keepaspectratio=true]{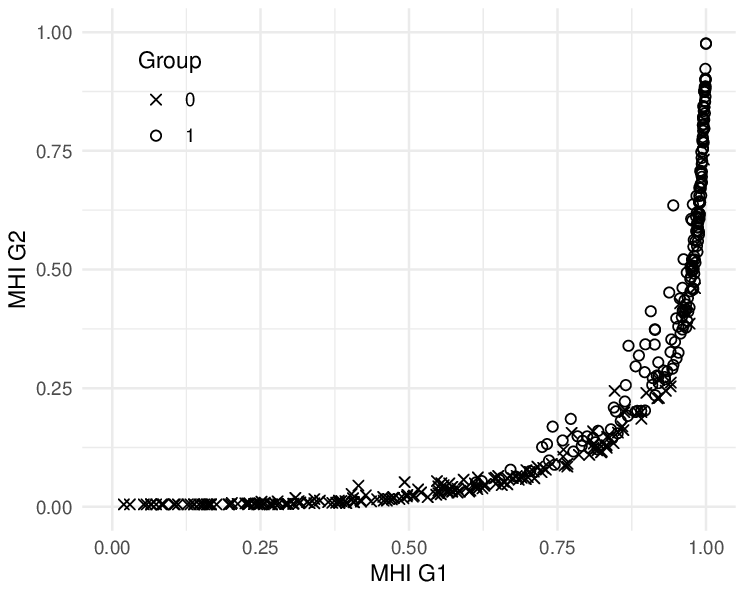}
  \caption{Experiment 6:\\ $EE_{H}$-plot}
\end{subfigure}
\caption{Illustration of EE-plots given by two distributions for different experiments}
\label{fig:2_experiments}
\end{figure}

In order to provide characterization for the EE-plots we observe the results derived from Figures \ref{fig:1_experiments} and \ref{fig:2_experiments}. In cases of similar distributions, a distinct 45-degree cluster line is evident. For sinusoidal distributions, a notable chromosome-like pattern emerges. As the amplitude widens or dispersion increases within the same chromosome shape, the data points become more scattered. Notably, distributions that do not overlap exhibit an L shape in the corners of the EE-plot. With slight overlapping, this L shape transforms into a more rounded form at the corners.

Therefore, as mentioned when two sample data follow the same distribution, data points cluster around the 45-degree line on the plot. This pattern is evident in Figures \ref{fig:1_experiments}(a), \ref{fig:1_experiments}(b), and \ref{fig:1_experiments}(c). However, as the distributions begin to diverge, points within the EE-plot start to disperse along the axis, leading to interesting visual patterns as demonstrated in \ref{fig:1_experiments}(d), \ref{fig:1_experiments}(e), and \ref{fig:1_experiments}(f). In instances where the curves do not overlap, a similar pattern emerges: data points within the EE-plot remain distinct, akin to what is depicted in Figures \ref{fig:1_experiments}(g), \ref{fig:1_experiments}(h), and \ref{fig:1_experiments}(i).

In order to provide the characterization of EE-plots, we analyze the outcomes depicted in Figures \ref{fig:1_experiments} and \ref{fig:2_experiments}. When distributions are similar, a clear 45-degree cluster line becomes evident, as observed in Figures \ref{fig:1_experiments}(a), \ref{fig:1_experiments}(b), and \ref{fig:1_experiments}(c). For sinusoidal overlapping distributions, presented in Experiments 2, 4 and 5 a distinctive chromosome-like pattern emerges, and as the amplitude or dispersion increases within this pattern, data points disperse, creating a scattered effect. 

The visual patterns in the EE-plots show transformations as distributions vary. In instances where curves do not overlap, a discernible L shape appears in the corners of the EE-plot, as illustrated in Figures \ref{fig:1_experiments}(g), \ref{fig:1_experiments}(h), and \ref{fig:1_experiments}(i). With slight overlapping, this L shape transitions into a more rounded form at the corners, introducing interesting visual nuances. These patterns offer valuable insights into the behavior of EE-plots, providing a visual means to interpret similarities and differences between distributions.

\subsection{EE-classifier}

The $\mathbb{R}^2$ space given by the EE-plot, provides patterns of how the groups are distributed. Therefore, as in \cite{bib9}, the groups can be separated by using traditional classification methods. A set of methods are proposed based on different techniques such as based on discriminant analysis (LDA, QDA), machine learning methods (SVM, Random Forest) and Nonparametric classification methods (kNN). One could implement different classification methods as there is no restriction on the data anymore to grasp their performance.

The EE-classifier works by taking two groups of data given by different distributions, measuring the modified epigraph or hypograph index values per curve for each distribution, creating the EE-plot and lastly inside the $\mathbb{R}^2$ space classify according the proposed classification methodologies. The results given by the different methods are measured in terms of accuracy as a metric. 

Note that the classifier works with both modified epigraph or hypograph indexes, as the EE-plot using one provides a mirrored version of the other. In this article, both indexes will be shown in order to demonstrate how the method is still useful when both of them are used.

\section{Applications}\label{sec4}
To evaluate the efficacy of the new EE-classifier we intend to assess the method using both synthetic data and real-world data sourced from R libraries. By comparing results and accuracy metrics for the classification, we aim to present an understanding of its capabilities. Moreover, we will include cross-validation accuracy metrics to demonstrate how the method performs across various data resampling, thereby providing confidence intervals for the metrics and outcomes.

\subsection{Synthetic Data}
In the synthetic data demonstration we use the data distributions given in both Figures \ref{fig:1_experiments} and \ref{fig:2_experiments}. The synthetic data was created with the \emph{generate\_gauss\_fdata} function from the \emph{roahd} package. This function generates a dataset of univariate functional data with a desired mean and covariance function. It takes as input the centerline of the distribution and a covariance operator. The centerline of the distribution is given by a set of functions and the covariance operator is created by the function \emph{exp\_cov\_function}, which takes in two different outputs $\alpha$ and $\beta$, they respectively correspond to the amplitude and dispersion of the parameters inside the exponential covariance function. Using different variations of this functions, six experiments were conducted in order to grasp the functionality of the EE-classifier, these are described in Table \ref{tab:experiments}.

The distributions generated for the experiments consist on 200 samples per group with 100 data points each. The construction of these experiments was based on how the EE-classifier would perform when using functional data with different characteristics. By using the same distribution (Experiment 1), different distribution with complementary functions, such as $sine$ ans $cosine$ (Experiment 2), non overlapping distributions (Experiment 3), distributions with wider amplitude (Experiment 4), distributions with higher dispersion (Experiment 5) and distributions with different functions as centerlines (Experiment 6). 

\begin{table}[h]
\footnotesize
\centering
\caption{Experiment and parameters definitions used for the illustration of the EE-classifier}\label{tab:experiments}
\tabcolsep=15.5pt
\begin{tabular}{|c|c|c|c|}
\hline \hline
Experiment & Centerline & $\alpha$ & $\beta$ \\
\hline
\multirow{2}*{1}  & $sin(\pi)$ & 0.2 & 0.3\\
& $sin(\pi)$ & 0.2 & 0.3 \\
\hline
\multirow{2}*{2}  & $sin(\pi)$ & 0.2 & 0.3\\
& $cos(\pi)$ & 0.2 & 0.3 \\
\hline
\multirow{2}*{3}  & $sin(\pi)-2$ & 0.2 & 0.3\\
& $sin(\pi)$ & 0.2 & 0.3 \\
\hline
\multirow{2}*{4}  & $sin(2\pi)$ & 0.2 & 0.3\\
& $cos(2\pi)$ & 0.7 & 0.3 \\
\hline
\multirow{2}*{5}  & $sin(2\pi)$ & 0.2 & 0.3\\
& $cos(2\pi)$ & 0.2 & 0.9 \\
\hline
\multirow{2}*{6}  & $x^2 - 1$ & 0.2 & 0.3\\
& $x^2$ & 0.2 & 0.9 \\
\hline \hline
\end{tabular}
\end{table}

\begin{figure}[h]
\centering

\begin{subfigure}{.45\textwidth}
  \centering
  \includegraphics[width=\linewidth, height=25cm,keepaspectratio=true]{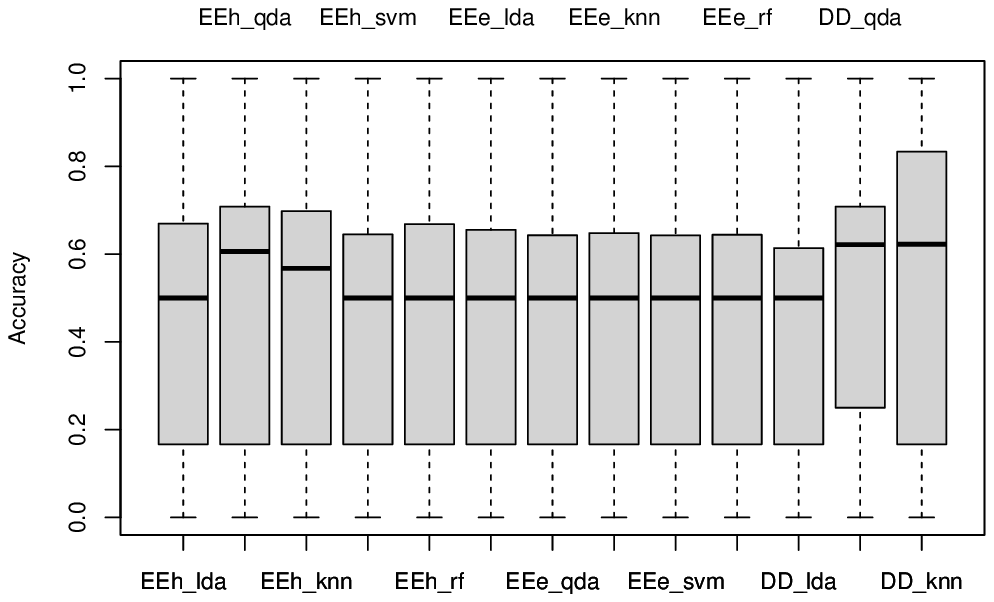}
  \caption{Experiment 1}
\end{subfigure}%
\begin{subfigure}{.45\textwidth}
  \centering
  \includegraphics[width=\linewidth, height=25cm,keepaspectratio=true]{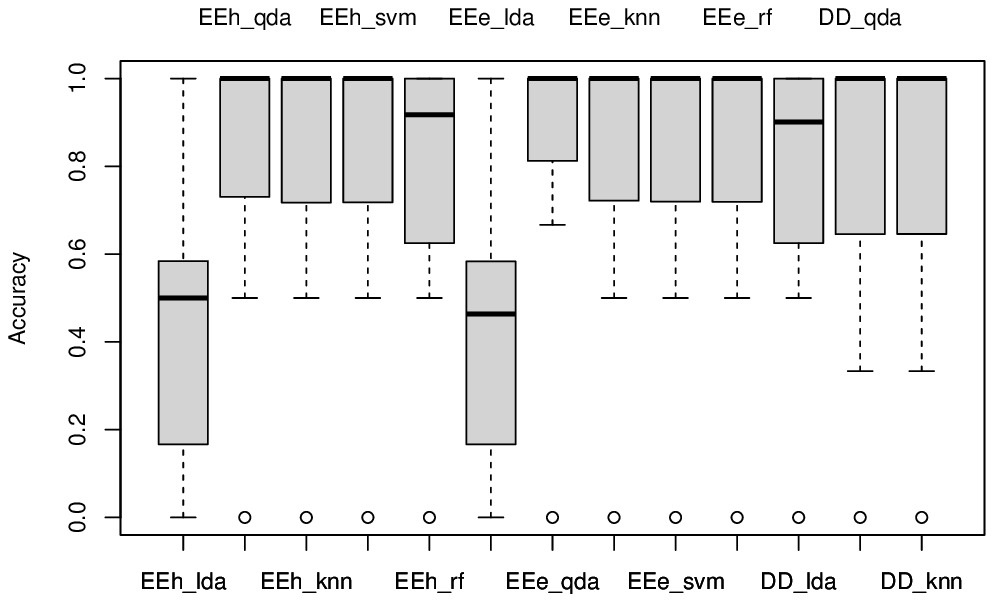}
  \caption{Experiment 2}
\end{subfigure}
\hfill
\begin{subfigure}{.45\textwidth}
  \centering
  \includegraphics[width=\linewidth, height=25cm,keepaspectratio=true]{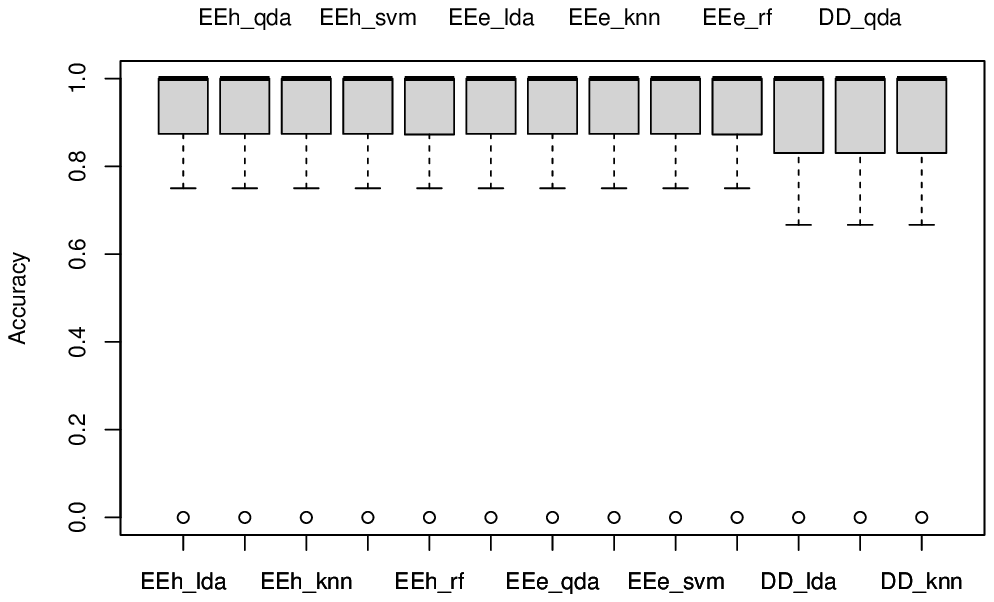}
  \caption{Experiment 3}
\end{subfigure}%
\begin{subfigure}{.45\textwidth}
  \centering
  \includegraphics[width=\linewidth, height=25cm,keepaspectratio=true]{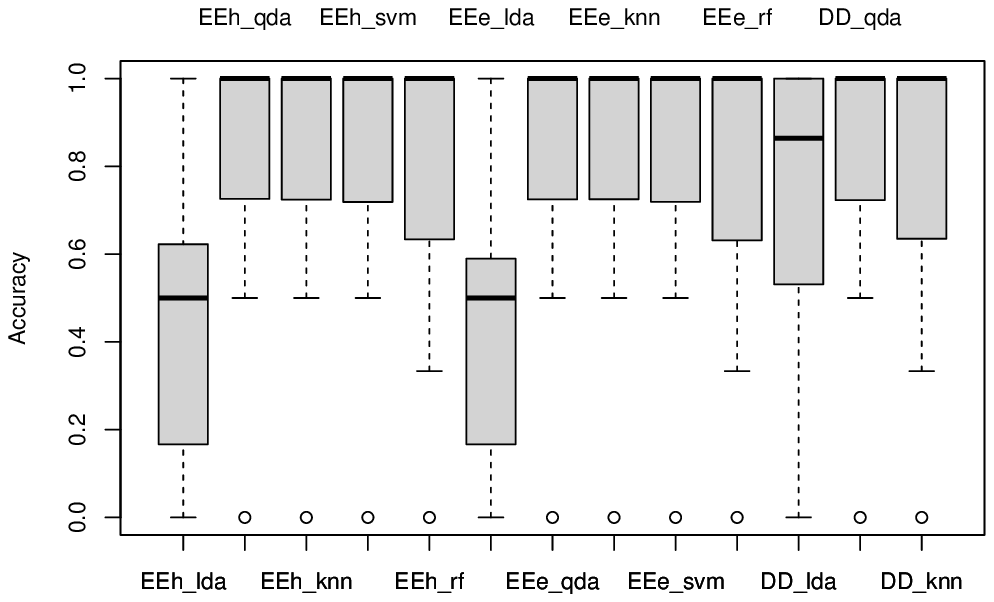}
  \caption{Experiment 4}
\end{subfigure}
\hfill
\begin{subfigure}{.45\textwidth}
  \centering
  \includegraphics[width=\linewidth, height=25cm,keepaspectratio=true]{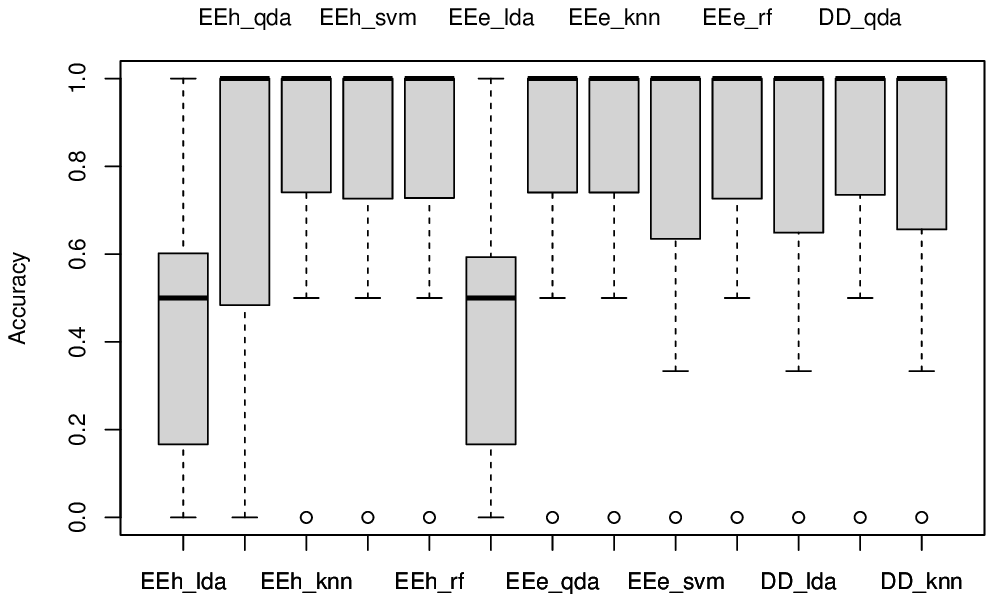}
  \caption{Experiment 5}
\end{subfigure}%
\begin{subfigure}{.45\textwidth}
  \centering
  \includegraphics[width=\linewidth, height=25cm,keepaspectratio=true]{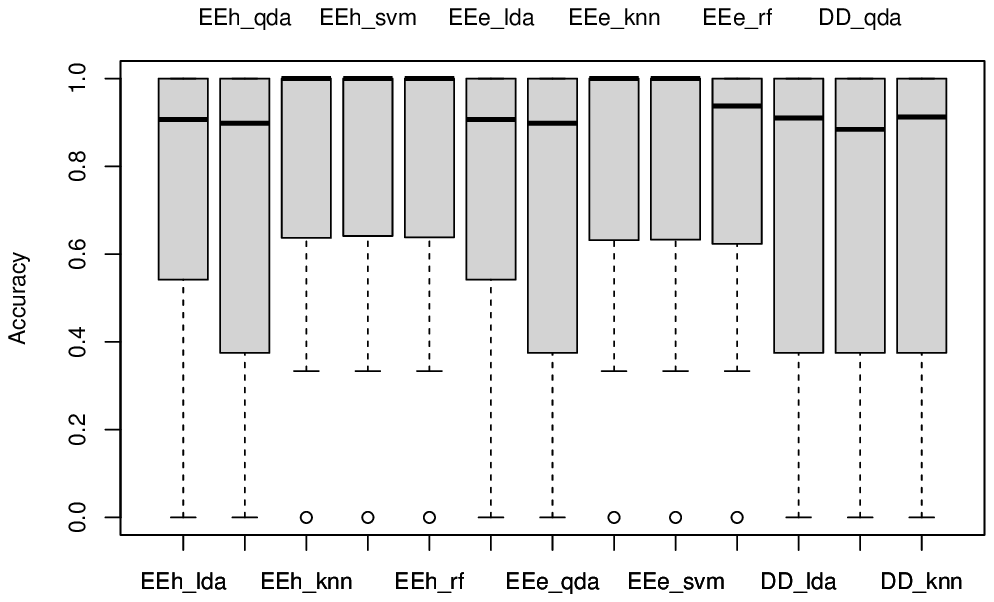}
  \caption{Experiment 6}
\end{subfigure}
\caption{Accuracy metrics given by each experiment and classifier using 100-
Fold cross-validation as train control}
\label{fig:results_synthetics}
\end{figure}

The EE-classifier was applied to these experiments, by using both the MEI and MHI for the EE-plot, and using the different classifiers (LDA, QDA, kNN, SVM and RF). To test the results we divide the sample into training (80\%) and testing (20\%) and train the methods using 100-Fold cross-validation as train control. 

To establish a benchmark, alongside the EE-classifier, we implemented the $DD^G$-classifier with the Fraiman \& Muniz depth in conjunction with several classifiers (LDA, QDA, and kNN). The selection of classifiers mirrors those chosen for the EE-classifier. This comparative analysis seeks to offer insights into the performance of the EE-classifier.

The results for each experiment are illustrated in Figure \ref{fig:results_synthetics}. In the first experiment, where the data is generated from the same distribution, all classifiers yield accuracy values around $\sim50\%$. This observation highlights their susceptibility to confusion when classifying closely positioned curves. Notably, the $DD^G$-classifier exhibits slightly superior performance in this scenario, but the EE-classifier emerges as a competitive alternative, offering smaller confidence intervals for accuracy metrics.

The LDA classifier, whether utilizing the $DD^G$-classifier or the EE-classifier, excels primarily when curves do not overlap, as demonstrated in the third experiment. This aligns with the characteristics of the LDA classifier, which relies on linear functions to segregate groups. Conversely, other classifiers (QDA, kNN, SVM, and RF) consistently achieve high classification accuracy values for both $EE_E$ and $EE_H$, with QDA generally providing the best overall results.

In the fourth and fifth experiments, where wider amplitude and higher dispersion are introduced, the classifiers' performance remains relatively consistent. This suggests that these variations do not significantly impact classifier efficacy. Interestingly, the EE-classifier demonstrates comparable results to the $DD^G$-classifier in these experiments.

\subsection{Real Data}

The evaluation of the EE-classifier on real data was performed with three different functional data sets which can all be found inside the R package \emph{fda.usc}. The first data set, \emph{Berkeley Growth Study} shows heights of 39 boys and 54 girls from age 1 to 18 and the ages at which they were collected. This data set was provided by \cite{bib16}, the research project focused on the physical development of a cohort of individuals from infancy to adulthood.

\begin{figure}[h]
\centering
\captionsetup[subfigure]{justification=centering}
\begin{subfigure}{.34\textwidth}
  \centering
  \includegraphics[width=.8\linewidth, height=10cm,keepaspectratio=true]{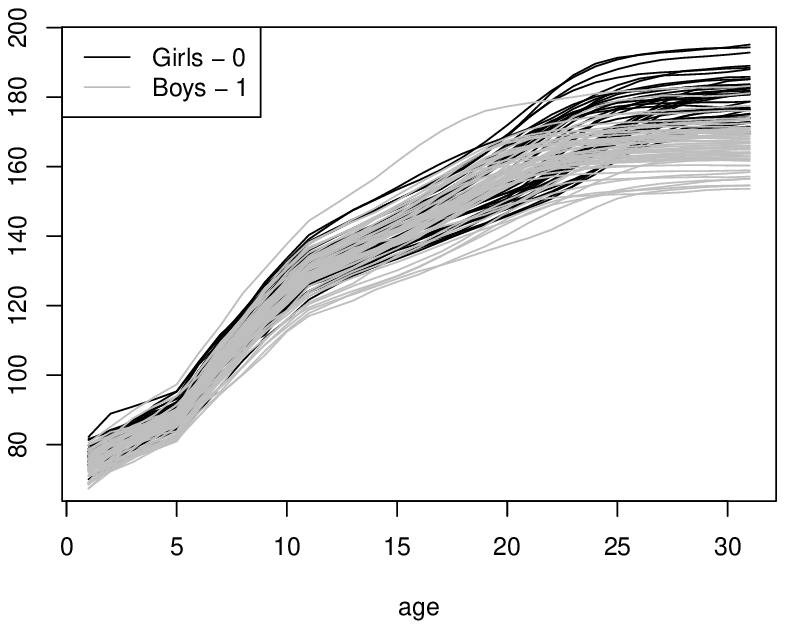}
  \caption{Height of boys and girls}
\end{subfigure}%
\begin{subfigure}{.33\textwidth}
  \centering
  \includegraphics[width=.8\linewidth, height=4cm,keepaspectratio=true]{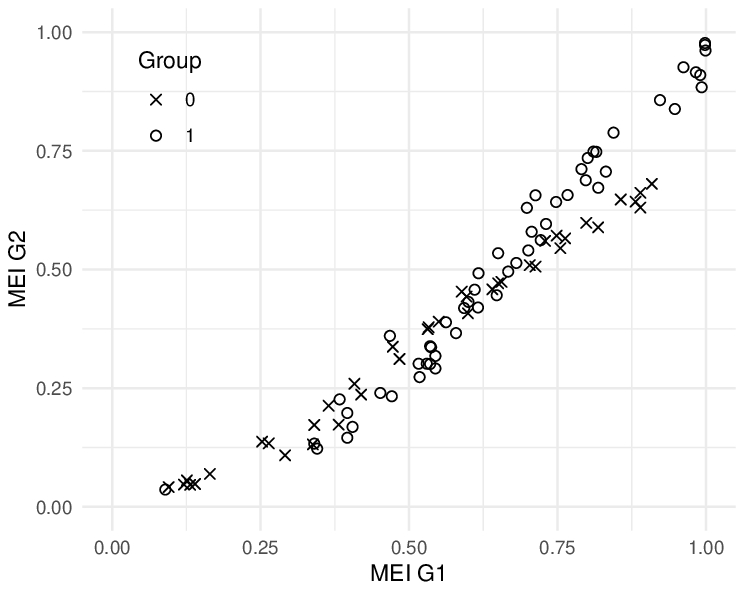}
  \caption{$EE_E$-plot}
\end{subfigure}%
\begin{subfigure}{.33\textwidth}
  \centering
  \includegraphics[width=.8\linewidth, height=4cm,keepaspectratio=true]{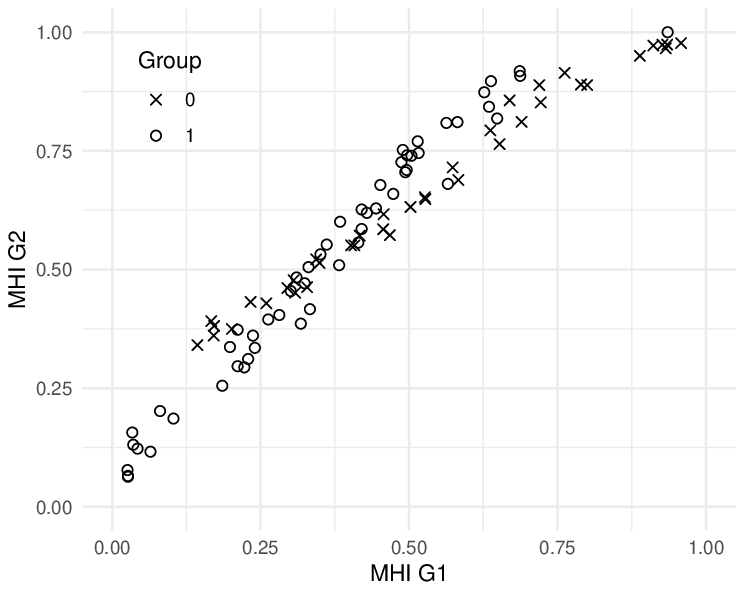}
  \caption{$EE_H$-plot}
\end{subfigure}
\caption{Curves and EE-plots of the Berkeley Growth Study data set}
\label{fig:growth}
\end{figure}

\begin{figure}[h]
    \centering
\includegraphics[width=0.6\textwidth,keepaspectratio=true]{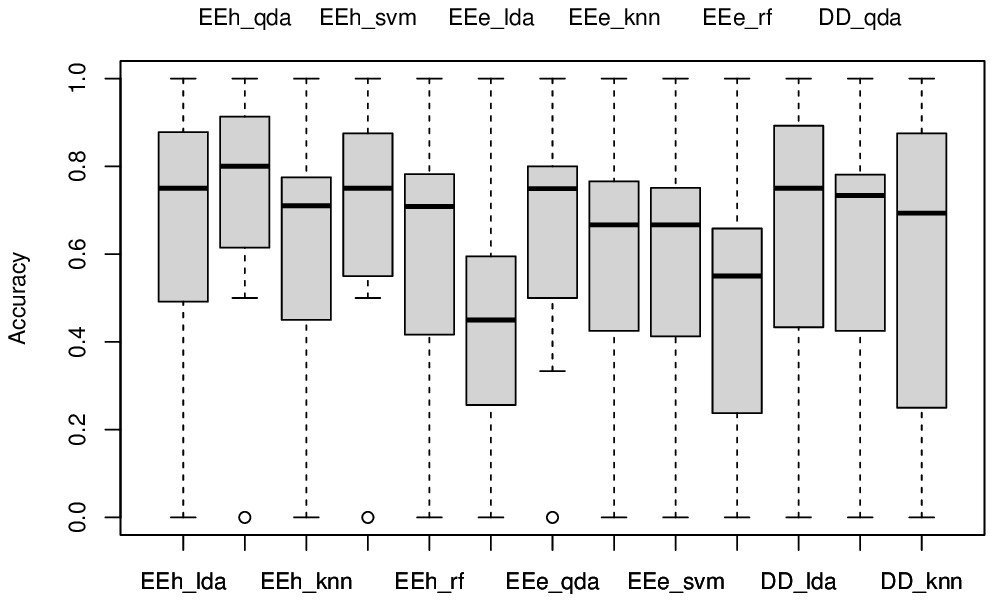}
    \caption{Berkeley Growth Study accuracy metrics given by each classifier using 20-Fold cross-validation as train control}
    \label{fig:results_growth}
\end{figure}

The curves for boys and girls are depicted in Figure \ref{fig:growth}(a), the $EE_E$-plot and $EE_H$-plot are shown in Figures \ref{fig:growth}(b) and  \ref{fig:growth}(c), respectively. We can see how the distributions for boys and girls are similar to each other, except for the oldest years where the girls overpass the height of the boys. Inside the the $EE_E$-plot and $EE_H$-plot we can also appreciate the similarity in the curves according the separation of both groups. The cross-validation results for this data set was done using 20-Folds and the results are given in Figure \ref{fig:results_growth}.

The distribution patterns for boys and girls show similarity, with the exception of the eldest years, where girls surpass boys in height. In both the $EE_E$-plot and $EE_H$-plot, the curves demonstrate similarity for the youngest years. However, in the oldest years, a noticeable separation of datapoints becomes evident, underscoring the separation between the two groups.

The cross-validation for this dataset utilized a 20-fold approach, and the results are depicted in Figure \ref{fig:results_growth}. To establish a benchmark, both the EE-classifier and the DD-classifier were implemented in the analysis.

The results reveal variations in accuracy metrics among the classifiers. Despite fluctuations, an overall accuracy of approximately $\sim75\%$ demonstrates the classifiers efficacy in separating between boys and girls based on height. Notably, the LDA classifier paired with the $EE_E$-plot shows the least favorable results, while the QDA-classifier along with the $EE_H$-plot emerges as the best performing combination.

In comparison with the $DD^G$-classifier, the EE-classifier demonstrates narrower confidence intervals and a smaller gap between the 1st and 3rd quartiles, indicating greater robustness. This suggests that the EE-classifier provides a more stable and reliable performance in this context.

\begin{figure}[h]
\centering
\captionsetup[subfigure]{justification=centering}
\begin{subfigure}{.34\textwidth}
  \centering
  \includegraphics[width=.8\linewidth, height=10cm,keepaspectratio=true]{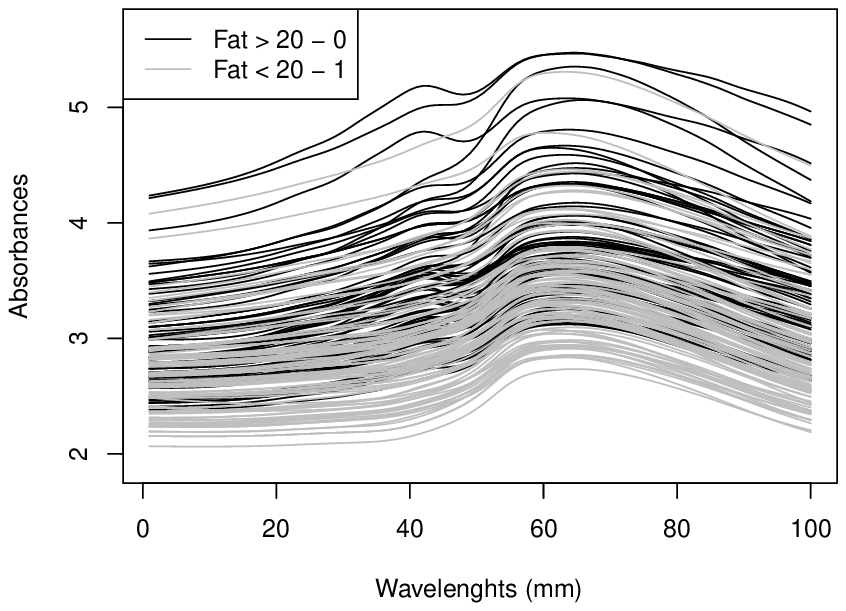}
  \caption{Meat curves}
\end{subfigure}%
\begin{subfigure}{.33\textwidth}
  \centering
  \includegraphics[width=.8\linewidth, height=4cm,keepaspectratio=true]{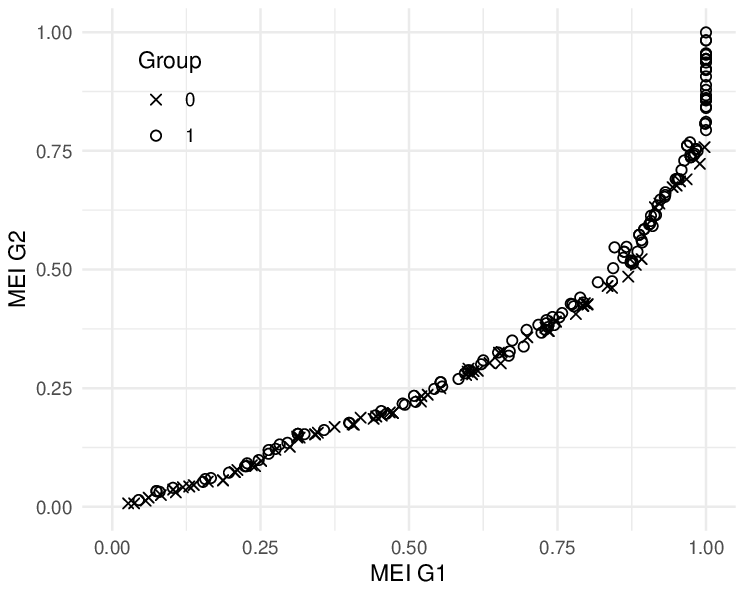}
  \caption{$EE_E$-plot}
\end{subfigure}%
\begin{subfigure}{.33\textwidth}
  \centering
  \includegraphics[width=.8\linewidth, height=4cm,keepaspectratio=true]{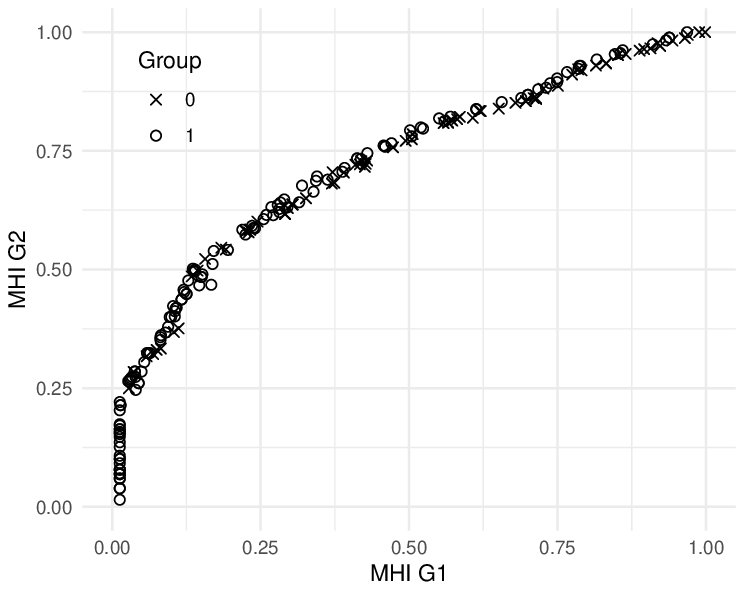}
  \caption{$EE_H$-plot}
\end{subfigure}
\caption{Curves and EE-plots of the Tecator data set}
\label{fig:tecator}
\end{figure}

The second dataset, referred to as \emph{Tecator}, comprises spectrometric curves representing absorbance measured at 100 different wavelengths for pieces of meat, with fat percentages categorized into two groups: small ($< 20\%$) and large ($> 20\%$) \cite{bib17}. The division of these curves into the respective fat percentage groups is illustrated in Figure \ref{fig:tecator}(a). Notably, absorbance levels tend to be higher for pieces with larger fat percentages. However, in the EE-plots (Figures \ref{fig:tecator}(b) and \ref{fig:tecator}(c)), it becomes evident that the distributions of absorbance values for the two fat percentage groups are intertwined. This close proximity of distributions suggests potential challenges for classifiers in separating them between the two classes.

\begin{figure}[h]
    \centering
\includegraphics[width=0.6\textwidth,keepaspectratio=true]{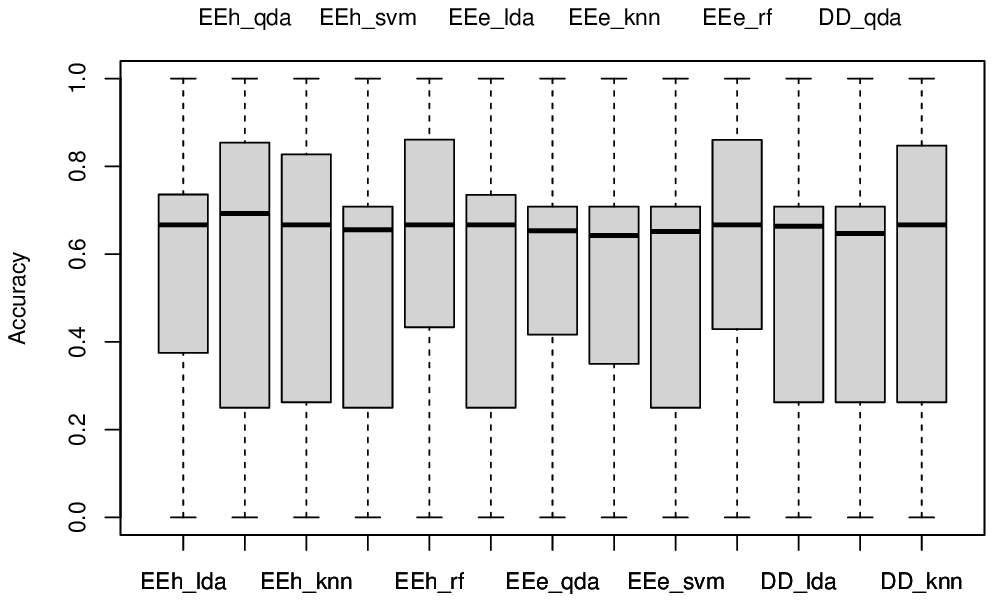}
    \caption{Tecator accuracy metrics given by each classifier using 50-Fold cross-validation as train control}
    \label{fig:results_tecator}
\end{figure}

The results presented in Figure \ref{fig:results_tecator} outline the accuracy metrics obtained from a cross-validation process employing 50-Folds. As intuitively anticipated, the classification rates are generally lower, reflecting the difficulty in separating the curves due to their close proximity. The accuracy metrics hover around $\sim70\%$, implying the capability to identify low-fat or high-fat curves at this classification rate. Despite the increased complexity of the dataset, the classifiers still demonstrate a reasonable level of effectiveness in distinguishing between the two categories.

The most favorable outcomes are given when employing the QDA classifier along the $EE_H$-plot, despite a relatively higher confidence interval. Remarkably, the EE-classifier consistently proves competitive in terms of results, showcasing notable robustness in certain instances by yielding smaller range between 1st and 3rd quartile. This shows the effectiveness of the EE-classifier in handling the challenging distributions such as the Tecator dataset.

\begin{figure}[h]
\centering
\captionsetup[subfigure]{justification=centering}
\begin{subfigure}{.34\textwidth}
  \centering
  \includegraphics[width=.8\linewidth, height=10cm,keepaspectratio=true]{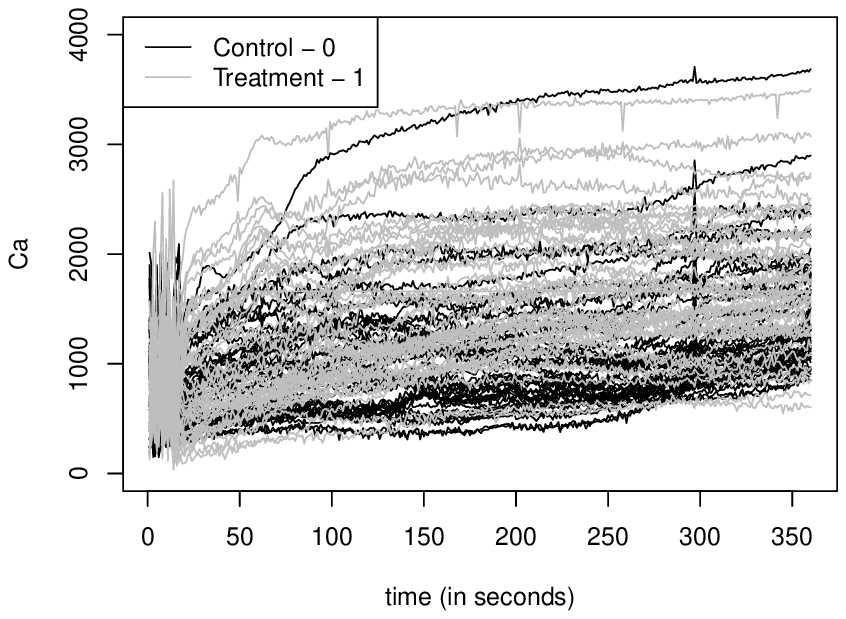}
  \caption{Control \& treatment curves}
\end{subfigure}%
\begin{subfigure}{.33\textwidth}
  \centering
  \includegraphics[width=.8\linewidth, height=4cm,keepaspectratio=true]{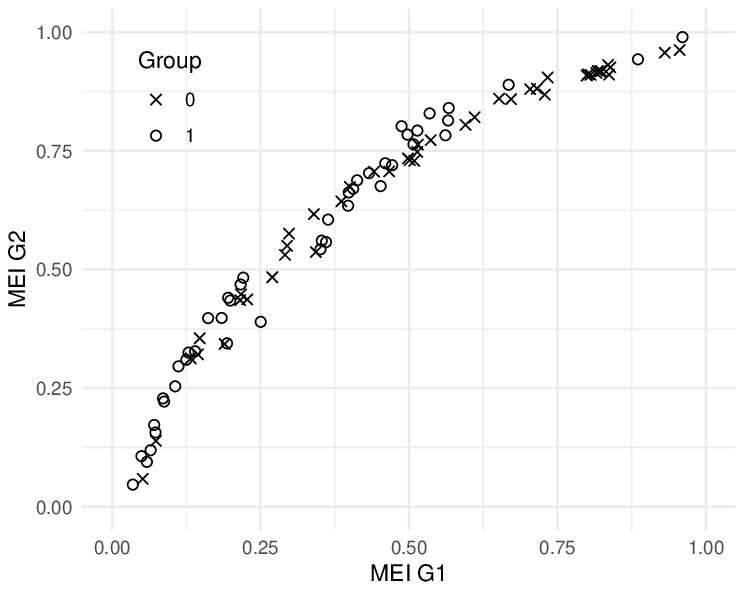}
  \caption{$EE_E$-plot}
\end{subfigure}%
\begin{subfigure}{.33\textwidth}
  \centering
  \includegraphics[width=.8\linewidth, height=4cm,keepaspectratio=true]{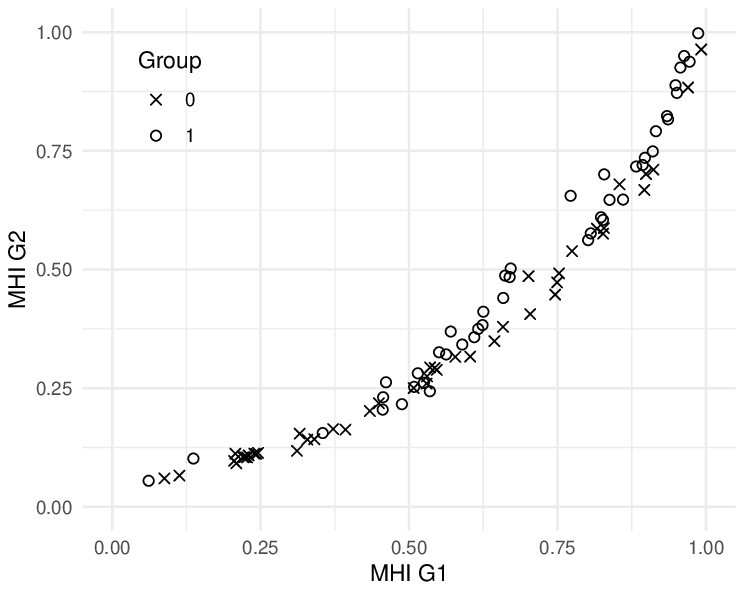}
  \caption{$EE_H$-plot}
\end{subfigure}
\caption{Curves and EE-plots of the MCO data set}
\label{fig:mco}
\end{figure}

The third and final dataset is the mithochondiral calcium overload (MCO) \cite{bib18}. This dataset measures the MCO in isolated mouse cardiac cells every 10 seconds during an hour which can be raised by drugs. It consists of two groups, one which receives no drug (control) and another group that receives the drug (treatment) that can raise MCO levels. Both groups and their MCO can be seen on Figure \ref{fig:mco}(a). The EE-plots for this dataset are shown on Figures \ref{fig:mco}(b) and \ref{fig:mco}(c) where the curves show slight differentiation of the groups.

\begin{figure}[h]
    \centering
\includegraphics[width=0.6\textwidth,keepaspectratio=true]{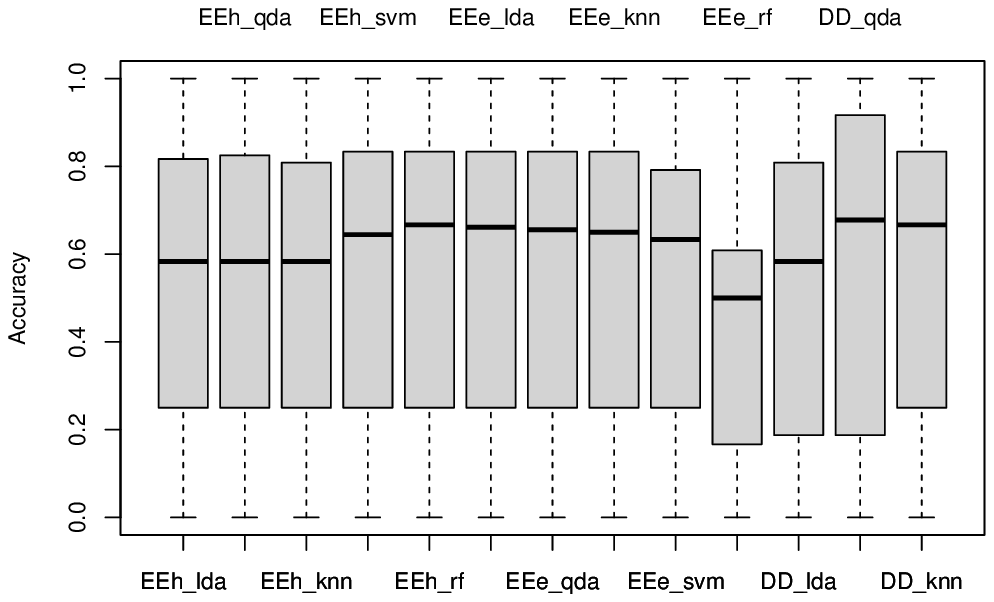}
    \caption{MCO accuracy metrics given by each classifier using 30-Fold cross-validation as train control}
    \label{fig:results_mco}
\end{figure}

The results for the MCO data set are shown in Figure \ref{fig:results_mco}. As we see in Figure \ref{fig:mco}(a) the curves are intertwined, and in Figures \ref{fig:mco}(b) and \ref{fig:mco}(c) the EE-plots show small distinction between the groups, therefore the results as shown in \ref{fig:results_mco} are not the best, with an average of correct classification of $\sim67\%$. 

The best overall result was achieved by using the $DD^G$-classifier with the QDA method. However, the range between the 1st and 3rd quartiles is high, indicating a wider variability in performance. In contrast, the EE-classifier using the $EE_E$-plot with QDA also shows good classification results, along with lower confidence intervals, showing enhanced robustness in comparison.

It is interesting to note that, despite the mean accuracy in some scenarios reflecting smaller classification rates, the cross-validation results often show a maximum accuracy of around $\sim100\%$ for most methods. This suggests that the EE-classifier is notably effective in classifying curves, even in scenarios where they have similar distributions. Furthermore, the consistent small ranges between the 1st and 3rd quartiles, as observed in both generated data and real datasets, underscore the robustness of the EE-classifier. This characteristic adds to its appeal as a reliable method capable of maintaining stable performance across various datasets, reinforcing its efficacy in practical applications.

\section{Case Study: S\&P 500 Fluctuation}\label{sec5}

The Standard and Poor's (S\&P) 500 stock value has been a huge focus between researchers, it aggregates the stock values for 500 different companies and provides an increase or decrease of the market for them combined in a single stock. The main challenge usually involves predicting a quantitative output value of the stock. Although, sometimes it is also desired to predict a categorical or qualitative output that shows whether the stock is increasing or decreasing \cite{bib19}. In this type of classification problem, predicting a numerical value gets substituted by predicting whether a given stock market performance will fall into the Up or Down category. Therefore, the goal is to find a model that can predict the direction in which the market will move in a certain period of time. 

\begin{figure}[h]
\centering
\captionsetup[subfigure]{justification=centering}
\begin{subfigure}{.34\textwidth}
  \centering
  \includegraphics[width=.8\linewidth, height=3cm]{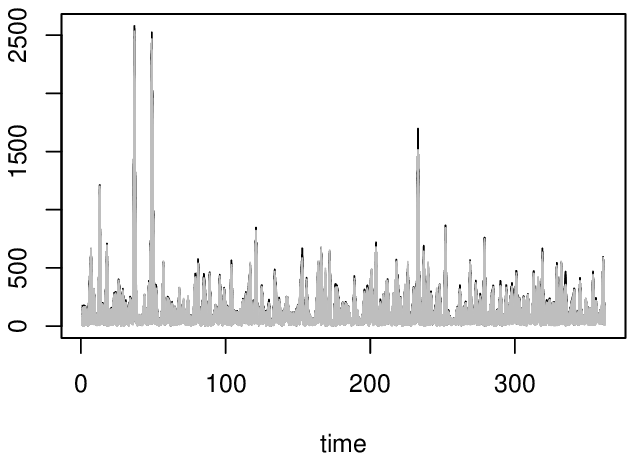}
  \caption{S\&P 500 stock price curves}
\end{subfigure}%
\begin{subfigure}{.33\textwidth}
  \centering
  \includegraphics[width=.8\linewidth, height=4cm,keepaspectratio=true]{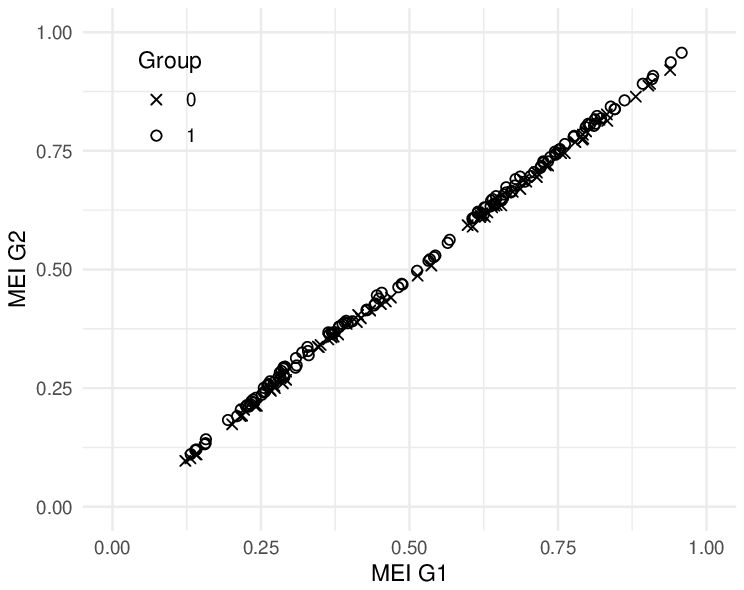}
  \caption{$EE_E$-plot}
\end{subfigure}%
\begin{subfigure}{.33\textwidth}
  \centering
  \includegraphics[width=.8\linewidth, height=4cm,keepaspectratio=true]{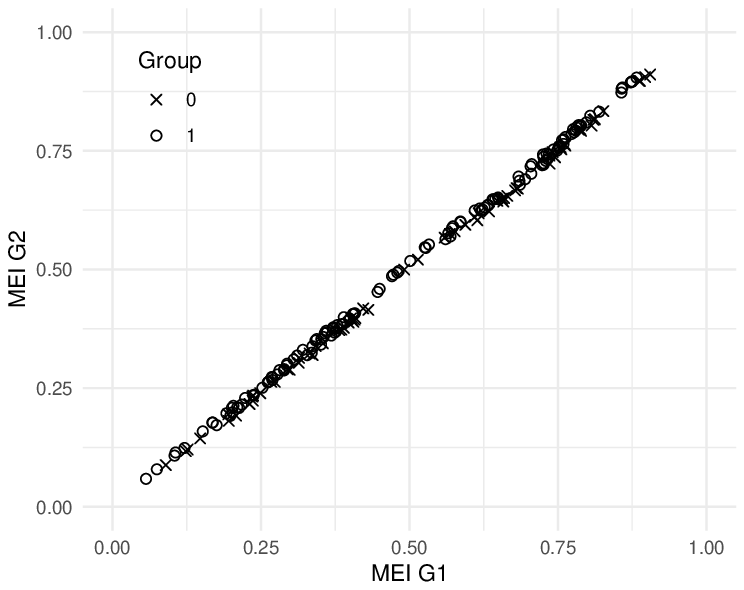}
  \caption{$EE_H$-plot}
\end{subfigure}
\caption{Curves and EE-plots of the  S\&P 500 stock price data set}
\label{fig:case}
\end{figure}

For this application, we received data for the stock values of 363 out of the 500 companies. We collected the closing stock values for each month from January 2005 to February 2023 for these companies. Subsequently, we created labels to indicate whether the S\&P 500 stock price went up or down compared to the previous month, corresponding to the stock values of these companies. The functional data array was then constructed with 218 observations, each containing 363 data points along with the corresponding labels.

\begin{figure}[h]
    \centering
\includegraphics[width=0.6\textwidth,keepaspectratio=true]{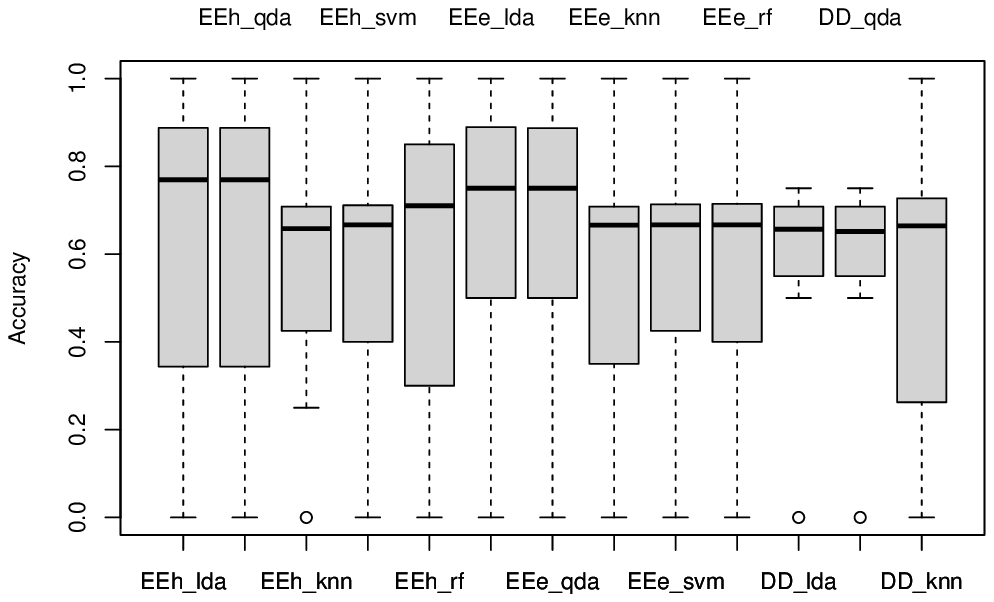}
    \caption{S\&P accuracy metrics given by each classifier using 50-Fold cross-validation as train control}
    \label{fig:results_case}
\end{figure}

The experiment was based on the example given in Section 4.7 found on page 171 of \cite{bib19} where the classification problem consists of the categorical prediction of the market increase or decrease. The data from January 2005 to December 2021 was used to predict if the market was going Up or Down for the dates between January 2022 to February 2023. Therefore, that data was separated as training and testing samples, on which we performed a 50-Fold cross-validation training and a prediction on the testing samples based on the trained models.

The S\&P curves are depicted in Figure \ref{fig:case}. These curves (Figure \ref{fig:case}(a)) exhibit considerable overlap, as evident in the EE-plot (Figure \ref{fig:case}(b), (c)). This overlap, as discussed in prior sections and supported by experimental findings, poses a potential challenge for the classifier.

The cross-validation results (Figure \ref{fig:results_case}) indicate that, on average, the models can predict whether the market will go Up or Down with an accuracy of approximately $\sim75\%$. Notably, the EE-classifier's best performance is observed with LDA and QDA, although with a wide confidence interval for both the $EE_E$-plot and $EE_H$-plot. In this context, the EE-classifier demonstrates higher mean accuracy metrics than the $DD^G$-classifier, while showing smaller ranges between the 1st and 3rd quartiles.

Regarding the prediction outcomes from the test sample, the accuracy metrics are detailed in Table \ref{tab:6_results}. Most of the metrics hover around $\sim50\%$, with the RF model using the $EE_E$-plot delivering the best result. These outcomes align with expectations, given the overlapping of datapoints evident in the EE-plots. However, it's worth noting that the DD-classifier also encounters challenges in classifying this data, emphasizing the complexity of the problem at hand.

Nonetheless, exploring the results of the classifiers, particularly the RF model using $EE_E$-plot, we see that out of 14 months on which the market increases the models correctly predicts 10 of them, and we anticipate a definite market increase for the years 2021-2023.

Despite the challenging nature of the dataset, a closer examination of the classifier results, especially the RF model using the $EE_E$-plot, one could say that, out of the 14 months in which the market shows an increase, the model correctly predicts 10 of them. This encouraging accuracy instills confidence, and we anticipate a market increase for the years 2022-2023 based on these predictive outcomes.

\begin{table}[h]
\footnotesize
\centering
\caption{Testing prediction results for the S\&P 500 data}\label{tab:6_results}
\tabcolsep=10pt 
\begin{tabular}{|l|c|c|c|}
\hline \hline
{Method} & {MEI} & {MHI} & DD \\
\hline 
LDA & 0.64         & 0.43    &   0.64  \\
QDA & 0.50         & 0.43        & 0.43\\
kNN & 0.35           & 0.40   & 0.57 \\
SVM & 0.45           & 0.47     & - \\
RF  & 0.71           & 0.69    & - \\
\hline 
\hline 
\end{tabular}
\end{table}

Even though the 2022-2023 data was not used to fit the training model, the level of accuracy of $\sim60\%$ is a good result for the S\&P 500 stock data, which can usually be hard to predict. This shows that the models act according to the relation between the companies stock value and fluctuation of the market. It is important also to note that the data set is quite small, and usually classification methods need larger data sets to be able to train and test appropriately. With larger history of the data, we could predict  S\&P 500 stock fluctuation more accurately.

\section{Conclusions}\label{sec6}
In this article, we immersed into supervised classification methods for functional data. Specifically, we focused on the depth-based classification methodology and introduced the EE-classifier as an extremality-based classifier. The classifier was examined using modified epigraph and modified hypograph indexes, showing the distinct behavior of various groups within the $\mathbb{R}^d$ space of the EE-plots.

Several experiments were carried out to demonstrate the effectiveness of the classifier using both synthetic and real data, alongside a benchmark provided by the $DD^G$-classifier. The classifier showed great results across various experiments, comparing different curves with distinct distribution characteristics, including variations in centerline, amplitude, and dispersion. Even in scenarios where curves are closely intertwined, making data classification slightly more challenging, the method exhibited good accuracy results.

Regarding classification methods, one might infer that the QDA classifier delivered superior outcomes in both training (cross-validation control) and testing. LDA proves to be among the top classifiers when dealing with curves from non overlapping distributions. As for the indexes, both MEI and MHI offer comparable accuracy metrics. However, in certain cases, the EE-classifier using MHI might shows better performance, making it an interesting question for further research and analysis.

The benchmark set by the $DD^G$-classifier serves as a reference point, providing valuable insights into the performance of the EE-classifier. Throughout the entirety of the article, the EE-classifier consistently showcased competitive results, suggesting its potential as a robust and effective method for classifying diverse types of distributions. This pattern underscores the EE-classifier's reliability and strength across various scenarios explored in this research.

Regarding the application to S\&P 500 stock values, the EE-classifier demonstrated a categorical prediction accuracy of approximately $\sim60\%$ for the fluctuations observed from January 2022 to February 2023. This outcome signifies the classifier's ability to correctly anticipate stock value increases during that period, aligning with the actual market trends between those months.

For both the real datasets and the application, while the obtained results were promising, it would be intriguing to conduct experiments with larger datasets. A more extensive dataset could offer additional information for training methods, potentially enhancing their performance. In an initial analysis of the application, only five years were considered. As we expanded the dataset to ten years, the metrics showed an upward trend, and concurrently, the confidence intervals for the mean accuracy in cross-validation diminished. This trend suggests the potential benefits of utilizing more extensive datasets for improved model training and more reliable performance assessments.

The classification method based on the extremality of functional data, gave a good performance and was able to classify the groups of every data set provided in each experiment and application in a positive light. Hopefully, this research leads to higher improvement of the method and new application can be found base on the EE-classifier. 

\section*{Acknowledgements}

The authors would like to thank the Finance Laboratory at Universidad EAFIT for providing the data for the S\&P 500 stock value. This research was founded by the project \emph {``Design of new statistical technics for the classification of high dimensional data"} with number 1119-11190072021 from Universidad EAFIT, Medellín, Colombia.

\bibliography{sn-bibliography}

\end{document}